# Catastrophic breakdown of the Caves model for quantum noise in some phase-insensitive linear amplifiers or attenuators based on atomic systems

Minchuan Zhou,[1] Zifan Zhou,[2] Selim M. Shahriar,[1,2,*]

[1]Department of Physics and Astronomy, Northwestern University, 2145 Sheridan Rd, Evanston, IL 60208, USA

[2]Department of EECS, Northwestern University, 2145 Sheridan Rd, Evanston, IL 60208, USA

*shahriar@northwestern.edu

When considering the effect of quantum noise (QN) in a phase-insensitive linear amplifier or attenuator, it is customary to use the single channel Caves model (SC-CM). Although this model is valid in simple situations, such as the presence of a beam splitter, it is not necessarily valid when a system with many degrees of freedom is involved. In order to address this issue, we consider in this paper various atomic transitions corresponding to amplification or attenuation using the master-equation- (ME-) based approach to model the QN and to compare the results with the SC-CM. For a four-level system that consists of a transition producing a broad gain peak and a transition producing an absorption dip, which results in perfect transparency at the center, we observe a catastrophic breakdown of the SC-CM. We also show that for a general two-level atomic system, the SC-CM does not apply, except in the limiting case when only either amplification or attenuation exists. A special case where the two models predict the same result is a Λ-type three-level electromagnetically induced transparency (EIT) system in which the QN at zero detuning vanishes while the system is in the dark state. We also study an optically pumped five-level gain EIT system which has a perfect transparency dip superimposed on a gain profile, and yields the negative dispersion suitable for use in enhancing the sensitivity-bandwidth product of an interferometric gravitational wave detector. In this case, we find that, for some set of parameters, the QN is vanishingly small at the center of the dip, and the SC-CM agrees closely with the ME model. However, we also find that for some other set of parameters, the SC-SM model disagrees strongly with the ME model. All these cases illustrate a wide range of variations in the degree of disagreement between the predictions of the SC-CM and the ME approaches.

## I. INTRODUCTION

The calculation of quantum noise (QN) is important for determining the fundamental limitations of various systems of interest in precision metrology. The quantum noise in a linear amplifier or attenuator can generally be determined by the canonical commutation relation, as given by the so-called single-channel Caves model (SC-CM) [1, 2]. The fundamental quantum limitation in a phase-insensitive linear amplifier has also been studied in terms of the average fidelity, which is experimentally testable [3].

Recently, we have proposed a gravitational wave detector that incorporates the white light cavity (WLC) [4, 5, 6, 7, 8, 9, 10] effect by adding a negative dispersion medium (NDM) in the signal recycling cavity in the advanced Laser Interferometer Gravitational Wave Observatory (aLIGO) [11, 12, 13, 14] design, which we call WLC-SR [15]. To calculate the QN for the detector, we need to consider the QN due to the NDM. Whereas the SC-CM predicts the level of QN for a given amplification or absorption factor, for an NDM realized using multiple atomic transitions, it is not *a priori* obvious whether the SC-CM is valid. In fact, it is easy to envision a case where the conclusion of the SC-CM runs counter to intuitive expectations. Consider, for example, a case where the medium consists of atoms of two different species. It is possible to prepare these two species in a way so that the probe will experience a relatively broad gain spectrum from one species and a narrower absorption spectrum from the other. By tuning parameters, such as the ratio of densities of the two species, it is possible to produce a net gain spectrum, which vanishes at a particular probe detuning, due to cancellation of nonzero gain from one species and the matching absorption from the other. At this detuning, the SC-CM predicts no QN. However, since each species has a significant population in the excited state at this condition, it seems obvious that there should be a considerable amount of quantum noise due to spontaneous emission from these atoms. In order to resolve this apparent inconsistency, it is necessary to determine the noise using a more fundamental approach, namely, the Markovian master equation (ME) [16] that describes the interaction among the atom, the semiclassical pump field, and the quantized probe mode. Here, we apply this approach to determine the QN for various types of phase-insensitive linear amplifiers or attenuators realized using atomic systems. We show the SC-CM predicts significantly different results compared to the ME model in some cases and identical results for some special cases. We do not find a general rule that can be applied to determine when the application of the SC-CM is expected to be a good approximation of the more exact result. As such, we conclude that one must always make use of the ME approach when dealing with resonant or near-resonant atomic systems. The technique presented in this paper would be useful in the treatment of quantum processes in linear phase-sensitive and phase-insensitive optical systems and would enable accurate evaluation of the QN in many systems of interest in precision metrology.

The rest of the paper is organized as follows. In Sec. II, we introduce the ME approach for calculating the QN in atomic systems. In Sec. III, we show the QN results in a two-level atomic system calculated using the ME, and then compare the results to the SC-CM. In Sec. IV, we apply the ME approach to a four-level atomic system, which produces a negative dispersion. In Sec. V, we show that in a $\Lambda$-type EIT electromagnetically induced transparency (EIT) [17, 18, 19, 20, 21, 22]) system, the results computed using the ME agree with the SC-CM, showing vanishing QN at zero detuning of the probe. Finally, in Sec. VI, we describe a scheme called gain-EIT (GEIT) that has the desired properties for enhancing the sensitivity-bandwidth product of the WLC-SR scheme [15] and analyze its QN. In Appendix A, we show the details of the steps for the ME approach for a two-level atomic system discussed in Sec. III. In Appendix B, we compare the susceptibilities of four different systems as determined by the ME model with the corresponding results obtained from semiclassical calculations, showing excellent agreement in each case.

## II. GENERAL OUTLINE OF THE MASTER EQUATION APPROACH

We consider first a general situation where a collection of non-interacting atoms is subjected to resonant or near-resonant optical fields. These fields consist of two parts: a pump field which is strong enough so that it can be described semiclassically and a single-mode probe field which is vanishingly small in intensity and is described quantum mechanically. (In some cases, the effect of the pump field may be modeled simply as a pumping rate from one state to another.) In addition, the atoms interact with a thermal reservoir of photons. For optical excitations, one can assume the temperature of the reservoir to be essentially zero so that the mean photon number in the thermal reservoir is zero. Under this condition, the effect of the thermal reservoir can be evaluated using the Weisskopf-Wigner theory of spontaneous emission [16]. The resulting evolution of the atomic system due to the interaction with the reservoir only can be modeled semiclassically by adding source, decay, and dephasing terms in the equation of motion for the (reduced) density matrix of the atoms. As such, the effect of the interaction with the reservoir modes does not need to be taken into account explicitly. If we define $\varrho_{a-f}$ as the density operator of the atom-field system, its evolution can be expressed as

$$\dot{\varrho}_{a-f} = -\frac{i}{\hbar}\left[\mathcal{H}, \varrho_{a-f}\right] + \dot{\varrho}_{a-f,R}, \tag{1}$$

where $\dot{\varrho}_{a-f,R}$ represents the source, decay, and dephasing terms resulting from the interaction with the reservoir and the Hamiltonian $\mathcal{H}$ is a sum of the atomic Hamiltonian $\mathcal{H}_A$, the field Hamiltonian $\mathcal{H}_F$ (excluding the reservoir), and the atom-field interaction Hamiltonian $\mathcal{H}_{AF}$ (again, excluding the interaction with the reservoir).

As we have shown in Ref. [23], the decay and dephasing of the reduced density operator for atoms can be accounted for by adding imaginary terms to the diagonal elements of $\mathcal{H}_A$, corresponding to half the decay rate of the corresponding atomic state. Even though here we are dealing with the atom-field density operator, this can still be performed, since $\mathcal{H}_A$ acts only on the atomic degree of freedom. For example, if an atomic state $|\alpha\rangle$ has a net decay rate of $\gamma_\alpha$, then the diagonal term of the atomic Hamiltonian for this state $\mathcal{H}_{\alpha\alpha}$ is changed to $\mathcal{H}_{\alpha\alpha} - i\gamma_\alpha/2$. When the net Hamiltonian is commuted with the density operator, this has the effect of adding a term, such as $\dot{\varrho}_{\alpha n,\alpha n'} = -\gamma_\alpha \varrho_{\alpha n,\alpha n'}$, where $|n\rangle$ and $|n'\rangle$ are the quantum states of the probe field. In addition, if a state $|\beta\rangle$ has a net decay rate of $\gamma_\beta$, then this change adds a term, such as $\dot{\varrho}_{\beta n,\beta n'} = -\gamma_\beta \varrho_{\beta n,\beta n'}$ as well as a dephasing term, such as $\dot{\varrho}_{\alpha n,\beta n'} = -[(\gamma_\alpha + \gamma_\beta)/2]\varrho_{\alpha n,\beta n'}$. Thus the quantum state of the probe field remains unaffected by the decay and dephasing caused by spontaneous emission. The source terms, accounting for the entry into certain states after they decay from higher-energy states, can be added explicitly [23]. For example, if atoms from level $|\alpha\rangle$ decay to level $|\beta\rangle$ at the rate of $\gamma_{\alpha\beta}$, then the source term would be of the form $\dot{\varrho}_{\beta n,\beta n'} = \gamma_{\alpha\beta}\varrho_{\alpha n,\alpha n'}$, where $|n\rangle$ and $|n'\rangle$ indicate the quantum states of the probe field. Thus the quantum state of the probe field remains conserved during the redistribution of population due to spontaneous emission.

Furthermore, we perform the rotating-wave approximation and then transform the system into an interaction picture. Denoting the interaction-picture density operator as $\rho_{a-f}$ and the interaction-picture Hamiltonian with complex diagonal elements as $\tilde{\mathcal{H}}'$, we can now write

$$\dot{\rho}_{a-f} = -\frac{i}{\hbar}\left(\tilde{\mathcal{H}}'\rho_{a-f} - \rho_{a-f}\tilde{\mathcal{H}}'^*\right) + \dot{\rho}_{a-f,Source}, \tag{2}$$

where $\dot{\rho}_{a-f,Source}$ represents the source terms. We can thus derive the equation of motion for the reduced density operator of the field $\tilde{\rho} \equiv Tr_{atom}(\rho_{a-f})$ following two steps: First, solve Eq. (2) for $\rho_{\alpha n,\beta n'} \equiv \langle \alpha, n|\rho_{a-f}|\beta, n'\rangle$, where $|\alpha\rangle$ and $|\beta\rangle$ are atomic states and $|n\rangle$ and $|n'\rangle$ are quantum states of the probe field; second, plug the solution into Eq. (2) and trace over all the atomic states to derive the equation of motion for the field density operator $\tilde{\rho}$. If $\tilde{\mathcal{H}}'$ were Hermitian, we would write

$$\dot{\tilde{\rho}} = -\frac{i}{\hbar} Tr_{atom}\left(\left[\tilde{\mathcal{H}}', \rho_{a-f}\right]\right). \tag{3}$$

However, since $\tilde{\mathcal{H}}'$ here is not Hermitian due to the addition of the complex terms, this becomes

$$\dot{\tilde{\rho}} = -\frac{i}{\hbar} Tr_{atom}\left(\tilde{\mathcal{H}}'\rho_{a-f} - \rho_{a-f}\tilde{\mathcal{H}}'^*\right). \tag{4}$$

Then we can derive the equation of motion for various moments of the annihilation ($a$) and creation ($a^\dagger$) operators of the field $a^{\dagger m}a^n$ by using the following relation:

$$\frac{d}{dt}\langle a^{\dagger m}a^n\rangle = Tr_{field}\left(a^{\dagger m}a^n \dot{\tilde{\rho}}\right). \tag{5}$$

Using Eq. (5), we can derive the QN, the details of which are described later.

### III. TWO-LEVEL ATOMIC SYSTEM

As a first example, let us consider the interaction of a closed two-level atom excited by a quantized probe with frequency $\upsilon$ as shown schematically in Fig. 1(a). The resonance frequency of the transition from the lower level $|b\rangle$ to the upper level $|a\rangle$ is $\omega = \omega_a - \omega_b$; the rate at which the atom is excited to $|a\rangle$ from $|b\rangle$ is $\gamma_{op}$; the rate of decay from $|a\rangle$ to $|b\rangle$ is $\gamma_a$. The optical pumping from $|b\rangle$ to $|a\rangle$ can in principle be achieved by coupling $|b\rangle$ to an auxiliary level $|c\rangle$ with a semiclassical laser field from which it could decay to an auxiliary level $|d\rangle$ and then to $|a\rangle$ as shown in Fig. 1(b). However, the net effect can be described by an incoherent pumping rate of $\gamma_{op}$.

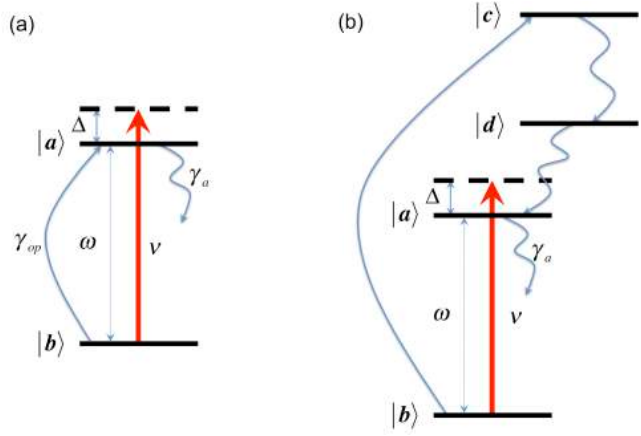

Fig. 1 (a) Two-level atomic system; (b) Schematic illustration of the optical pumping $\gamma_{op}$ from level $|b\rangle$ to $|a\rangle$.

Let us denote the annihilation and creation operators of the field by $a$ and $a^\dagger$. The Hamiltonian after rotating-wave transformation and adding the complex terms is then

$$\tilde{\mathcal{H}}' = \tilde{\mathcal{H}} - \frac{i}{2}\hbar\gamma_a|a\rangle\langle a| - \frac{i}{2}\hbar\gamma_{op}|b\rangle\langle b|. \tag{6}$$

where $g = -\mathcal{P}\mathcal{E}/\hbar$ is the coupling constant for the transition $|a\rangle \to |b\rangle$, $\mathcal{P}$ is the electric dipole moment, and $\mathcal{E} = (2\pi\hbar\nu/V)^{1/2}$. Note here that, for simplicity, we assume that the atoms are sitting inside a unidirectional ring cavity and $V$ is the volume of the cavity mode. Therefore the equation of motion can be written as

$$\dot{\rho}_{a-f} = -\frac{i}{\hbar}\left(\tilde{\mathcal{H}}'\rho_{a-f} - \rho_{a-f}\tilde{\mathcal{H}}'^*\right) + \dot{\rho}_{a-f,Source}. \tag{7}$$

The explicit set of equations for the density matrix elements are shown in Appendix A as Eqs. (A10)–(A13). It contains an infinite number of equations since the value of $\{n,n'\}$ extends from zero to infinity as illustrated in Fig. 2 where the couplings between the matrix elements are indicated by the curves with arrows. We can see that all the couplings are within a given manifold (column), except for the couplings shown by the slanted and dashed lines, which couple elements in different manifolds. Here, we have indexed, arbitrarily, each manifold by the unprimed number of photons accompanying atom in the $|a\rangle$ state. We will follow this indexing convention throughout this paper.

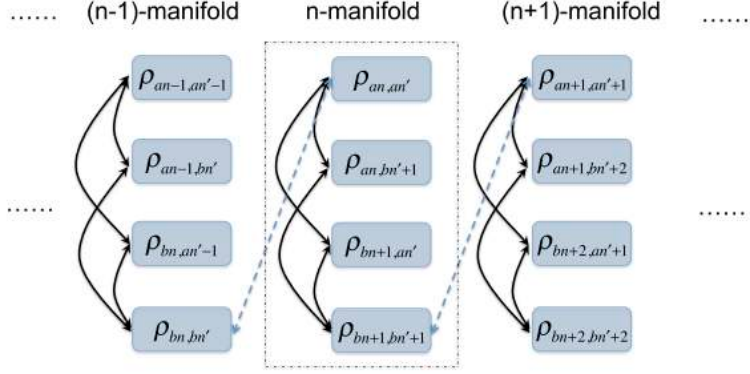

Fig. 2 Illustration of atom-field density matrix elements and their couplings.

Using the relation that $\rho_{an,an'} + \rho_{bn,bn'} = \tilde{\rho}_{nn'}$ and $\rho_{an+1,an'+1} + \rho_{bn+1,bn'+1} = \tilde{\rho}_{n+1,n'+1}$, we can decouple different manifolds. Then the couplings between different manifolds are removed, and we get a closed system $\rho_{an,an'}$, $\rho_{an,bn'+1}$, $\rho_{bn+1,an'}$, $\rho_{bn+1,bn'+1}$ (enclosed by a dashed box in Fig. 3) within the infinite set of density-matrix elements, described by Eqs. (A11), (A12), (A14), and (A15) in Appendix A, and coupled to source terms $\tilde{\rho}_{nn'}$, $\tilde{\rho}_{n+1,n'+1}$, where we recall that $\tilde{\rho}$ is the reduced density operator for the field only. The solution of these equations in the linear regime where $g$ is very small (shown in Appendix A) are plugged into the equation of motion for the reduced density matrix of the field as in Eq. (4), which can be written as

$$\dot{\tilde{\rho}}_{nn'} = -\frac{i}{\hbar}\left( \tilde{\mathcal{H}}'_{an,bn+1} \rho_{bn+1,an'} - \rho_{an,bn'+1} \tilde{\mathcal{H}}'^{*}_{bn'+1,an'} + \tilde{\mathcal{H}}'_{bn,an-1} \rho_{an-1,bn'} - \rho_{bn,an'-1} \tilde{\mathcal{H}}'^{*}_{an'-1,bn'} \right). \tag{8}$$

In the following subsections we discuss the results under resonant and nonresonant conditions.

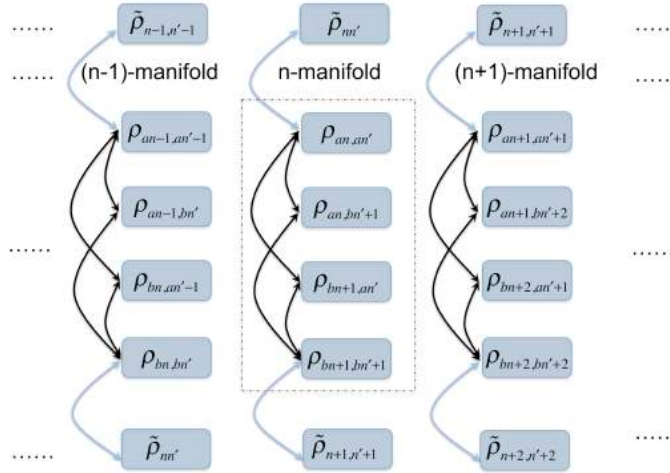

Fig. 3 Illustration of atom-field density-matrix elements and their coupling after removing the coupling between different manifolds.

### A. Resonant case

In the resonant case $\Delta = 0$, we have

$$\dot{\tilde{\rho}}_{nn'} = -\mathcal{A}_{res}\left[(n+n'+2)\tilde{\rho}_{nn'} - 2\sqrt{nn'}\tilde{\rho}_{n-1,n'-1}\right] - \mathcal{B}_{res}\left[(n+n')\tilde{\rho}_{nn'} - 2\sqrt{(n+1)(n'+1)}\tilde{\rho}_{n+1,n'+1}\right], \quad (9)$$

where $\mathcal{A}_{res} = g^2\gamma_{op}/(2\gamma_{ab}^2), \mathcal{B}_{res} = g^2\gamma_a/(2\gamma_{ab}^2)$. Equation (9) can also be written as

$$\dot{\tilde{\rho}} = -\mathcal{A}_{res}(aa^\dagger\tilde{\rho} - 2a^\dagger\tilde{\rho}a + \tilde{\rho}aa^\dagger) - \mathcal{B}_{res}(a^\dagger a\tilde{\rho} - 2a\tilde{\rho}a^\dagger + \tilde{\rho}a^\dagger a), \quad (10)$$

To prove the equivalence, we show that since $a|n\rangle = \sqrt{n}|n-1\rangle, a^\dagger|n\rangle = \sqrt{n+1}|n+1\rangle$, we get

$$\langle n|aa^\dagger\tilde{\rho}|n'\rangle = (n+1)\tilde{\rho}_{nn'}, \quad \langle n|a^\dagger\tilde{\rho}a|n'\rangle = \sqrt{nn'}\tilde{\rho}_{n-1,n'-1}, \quad \langle n|\tilde{\rho}aa^\dagger|n'\rangle = (n'+1)\tilde{\rho}_{nn'}, \quad (11)$$

$$\langle n|a^\dagger a\tilde{\rho}|n'\rangle = n\tilde{\rho}_{nn'}, \quad \langle n|a\tilde{\rho}a^\dagger|n'\rangle = \sqrt{(n+1)(n'+1)}\tilde{\rho}_{n+1,n'+1}, \quad \langle n|\tilde{\rho}a^\dagger a|n'\rangle = n'\tilde{\rho}_{nn'}. \quad (12)$$

We can derive the equations of motion for various moments of $a$ and $a^\dagger$ from Eq. (10) using Eq. (5),

$$\frac{d}{dt}\langle a\rangle = (\mathcal{A}_{res} - \mathcal{B}_{res})\langle a\rangle, \quad (13)$$

$$\frac{d}{dt}\langle a^\dagger a\rangle = 2(\mathcal{A}_{res} - \mathcal{B}_{res})\langle a^\dagger a\rangle + 2\mathcal{A}_{res}, \quad (14)$$

$$\frac{d}{dt}\langle a^2\rangle = 2(\mathcal{A}_{res} - \mathcal{B}_{res})\langle a^2\rangle. \quad (15)$$

Therefore the moments of $a$ and $a^\dagger$ as functions of time are determined as follows:

$$\langle a\rangle_t = \sqrt{G_{res}}\langle a\rangle_0 \quad (16)$$

$$\langle a^\dagger a\rangle_t = G_{res}\langle a^\dagger a\rangle_0 + (G_{res}-1)\frac{\mathcal{A}_{res}}{\mathcal{A}_{res}-\mathcal{B}_{res}} \quad (17)$$

$$\langle a^2\rangle_t = G_{res}\langle a^2\rangle_0 \quad (18)$$

where $G_{res} = \exp[2(\mathcal{A}_{res} - \mathcal{B}_{res})t]$. If we define two quadratures as $X_\theta$ and $X_{\theta+\pi/2}$ with

$$X_\theta = \frac{1}{2}(a^\dagger e^{i\theta} + ae^{-i\theta}), \quad (19)$$

it can be shown that

$$\langle X_\theta\rangle_t = \sqrt{G_{res}}\langle X_\theta\rangle_0, \quad (20)$$

$$\langle \Delta X_\theta^2\rangle_t = G_{res}\langle \Delta X_\theta^2\rangle_0 + (G_{res}-1)\frac{\mathcal{A}_{res}+\mathcal{B}_{res}}{4(\mathcal{A}_{res}-\mathcal{B}_{res})}. \quad (21)$$

If we write the evolution of the system as

$$a_t = \sqrt{G_{res}}\, a_0 + X_1 F_1^\dagger + X_2 F_2,  \tag{22}$$

$$X_1 = \sqrt{(G_{res}-1)\frac{\mathcal{A}_{res}}{\mathcal{A}_{res}-\mathcal{B}_{res}}}, \quad X_2 = \sqrt{(G_{res}-1)\frac{\mathcal{B}_{res}}{\mathcal{A}_{res}-\mathcal{B}_{res}}},  \tag{23}$$

where $a_0$ is the input, $a_t$ is the output, $F_1^\dagger$ and $F_2$ are vacuum modes, and $[F_i, F_j^\dagger] = \delta_{ij}$ $(i,j=1,2)$, then it will give us a power gain of $G_{res}$ and the same noise as that in Eq. (21). The commutation relation that $[a_k, a_k^\dagger] = 1, a_k = a_0, a_t$ will also be preserved.

We next compare these results to the Caves model. Let us first recall briefly the SC-CM [1, 2]. For a phase-insensitive linear amplifier, the SC-CM can be written as

$$a_t = \sqrt{G}\, a_0 + \sqrt{G-1}\, F^\dagger, \quad G > 1.  \tag{24}$$

This is illustrated in Fig. 4(a) where the amplifier is modeled as a beam combiner with two inputs and an output. Here, $G$ is the power gain, and $F^\dagger$ is the vacuum mode, which is responsible for the added noise. Note here that the commutation relation for the output is preserved since $[F, F^\dagger] = 1$. We can get

$$\langle X_\theta \rangle_t = \sqrt{G}\, \langle X_\theta \rangle_0,  \tag{25}$$

$$\langle \Delta X_\theta^2 \rangle_t = G \langle \Delta X_\theta^2 \rangle_0 + \frac{1}{4}(G-1).  \tag{26}$$

On the other hand, for a phase-insensitive linear attenuator, the SC-CM can be written as

$$a_t = \sqrt{G}\, a_0 + \sqrt{1-G}\, F, \quad G < 1,  \tag{27}$$

where $G$ is the power attenuation, and the vacuum mode is written as $F$ in order to preserve the commutation relation. This is illustrated in Fig. 4(b). Therefore, we get Eq. (25) and

$$\langle \Delta X_\theta^2 \rangle_t = G \langle \Delta X_\theta^2 \rangle_0 + \frac{1}{4}(1-G).  \tag{28}$$

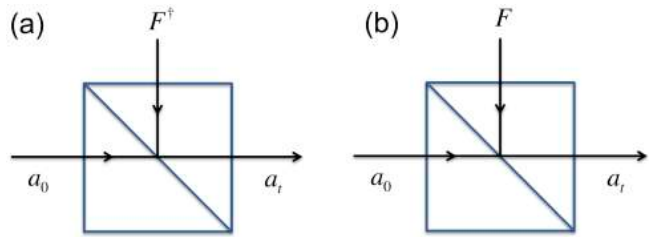

Fig. 4 Illustration of the SC-CM for (a) a phase-insensitive linear amplifier and (b) a phase-insensitive linear attenuator.

It can be seen that in the two-level atomic system the evolution of the mean value for the quadratures using the ME approach agrees with the result of the SC-CM, whereas the results for evolution of the variance differ. However, in the limiting case with pure amplification where $\gamma_{op} \neq 0$ and $\gamma_a = 0$, we have $\mathcal{A}_{res} = g^2/(2\gamma_{op})$ and $\mathcal{B}_{res} = 0$, then $G_{res} = \exp[2\mathcal{A}_{res}t]$, $X_1 = \sqrt{G_{res}-1}$, and $X_2 = 0$, thus arriving at the same results as in Eqs. (24)–(26). Similarly, in the limiting case with pure attenuation where $\gamma_a \neq 0$ and $\gamma_{op} = 0$, we have $\mathcal{A}_{res} = 0$ and $\mathcal{B}_{res} = g^2/(2\gamma_a)$, then $G_{res} = \exp[-2\mathcal{B}_{res}t]$, $X_2 = \sqrt{1-G_{res}}$, and $X_1 = 0$, therefore we arrive at the same results as in Eqs. (27) and (28). To summarize, a general two-level atomic system cannot be described by the SC-CM. For a two-level system, the SC-CM applies only when pure amplification or attenuation exists.

Since the single-channel Caves model does not agree with the ME model in a general two-level system, we may attempt to build a two-channel Caves model. Extending the beam combiner approach for the SC-CM model, we now use two beam combiners. As shown in Fig. 5(a), the left beam combiner introduces an amplification ($G_1$) whereas the right one introduces an attenuation ($G_2$), and there is also an input vacuum noise for each beam combiner: $F_1^\dagger$ and $F_2$. The output of the left and the right beam combiners are determined by

$$\tilde{a}_t = \sqrt{G_1}a_0 + \sqrt{G_1-1}F_1^\dagger, \tag{29}$$

$$a_t = \sqrt{G_2}\tilde{a}_t + \sqrt{1-G_2}F_2, \tag{30}$$

For the output of each beam combiner, the commutation relation is preserved. In order to get the correct gain factor and the correct amount of noise for the quadratures of the field as shown in Eqs. (20) and (21), we can choose the attenuation factor for the second channel and the amplification factor for the first channel respectively as $G_2 = 1-(G_{res}-1)\mathcal{B}_{res}/(\mathcal{A}_{res}-\mathcal{B}_{res})$ and $G_1 = G_{res}/G_2$.

However, there are two problems with this model. First, in order for a two-channel Caves model to be a useful construct, one must be able to infer the values of $G_1$ and $G_2$ via mere inspection of the system. As we see here, this is not at all the case. One must construct the ME model first in order to determine what $G_1$ and $G_2$ have to be. Second, and more importantly, even the two-channel Caves model where $G_1$ and $G_2$ are determined from the ME gives the correct value of QN only when the probe is propagating from left to right. To see this, consider the situation where the probe propagates from right to left as illustrated in Fig. 5(b). In this case, the outgoing field $a'_t$ will not be the same as $a_t$: The amplitude of the net gain for the two stages is again $G_{res}$, but the additional noise is

$$\langle \Delta X_\theta^2 \rangle_{noise} = (G_{res} - 1)\frac{\mathcal{A}_{res} + \mathcal{B}_{res}G_{res}}{4(\mathcal{A}_{res} - \mathcal{B}_{res}G_{res})}, \tag{31}$$

which is different from that in Eq. (21). Therefore, it is not possible to construct a two-channel Caves model represented in terms of effective beam combiners.

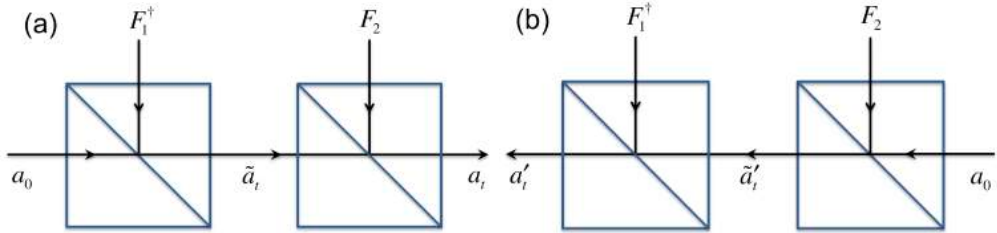

Fig. 5 (a) Illustration of two-channel Caves model and (b) illustration of oppositely propagating fields in the two-channel Caves model.

## B. Nonresonant case

When the field is off resonance with the two-level atom ($\Delta \neq 0$), the equation of motion for the field is

$$\dot{\tilde{\rho}}_{nn'} = -\mathcal{A}\left[(n+n'+2)\tilde{\rho}_{nn'} - 2\sqrt{nn'}\tilde{\rho}_{n-1,n'-1}\right] - \mathcal{B}\left[(n+n')\tilde{\rho}_{nn'} - 2\sqrt{(n+1)(n'+1)}\tilde{\rho}_{n+1,n'+1}\right] \\ + i\mathcal{N}(n-n')\tilde{\rho}_{nn'}, \tag{32}$$

where

$$\mathcal{A} = \frac{g^2\gamma_{op}}{2(\gamma_{ab}^2 + \Delta^2)}, \quad \mathcal{B} = \frac{g^2\gamma_a}{2(\gamma_{ab}^2 + \Delta^2)}, \quad \mathcal{N} = \frac{g^2\Delta}{\gamma_{ab}^2 + \Delta^2}\cdot\frac{\gamma_{op} - \gamma_a}{2\gamma_{ab}}. \tag{33}$$

Equation (32) can be written in the form

$$\dot{\tilde{\rho}} = -\mathcal{A}(aa^\dagger\tilde{\rho} - 2a^\dagger\tilde{\rho}a + \tilde{\rho}aa^\dagger) - \mathcal{B}(a^\dagger a\tilde{\rho} - 2a\tilde{\rho}a^\dagger + \tilde{\rho}a^\dagger a) + i\mathcal{N}\left(a^\dagger a\tilde{\rho} - \tilde{\rho}a^\dagger a\right). \tag{34}$$

Note that the values of $\mathcal{A}$ and $\mathcal{B}$ are different compared to Eq. (10). Furthermore, there is an additional term $i\mathcal{N}\left(a^\dagger a\tilde{\rho} - \tilde{\rho}a^\dagger a\right)$, resulting in a phase shift in the output, as will be shown later. When $\Delta = 0$, we have $\mathcal{A} = \mathcal{A}_{res}, \mathcal{B} = \mathcal{B}_{res}$, and $\mathcal{N} = 0$, and the results reduce to the same as those in Sec. III A. We have shown in Appendix B 1 that the susceptibility of this system determined by the ME model agrees with the result following the semiclassical approach.

We can now derive the equations of motion for moments of $a$ and $a^\dagger$,

$$\frac{d}{dt}\langle a \rangle = (\mathcal{A} - \mathcal{B} + i\mathcal{N})\langle a \rangle, \tag{35}$$

$$\frac{d}{dt}\langle a^\dagger a \rangle = 2(\mathcal{A} - \mathcal{B})\langle a^\dagger a \rangle + 2\mathcal{A}, \tag{36}$$

$$\frac{d}{dt}\langle a^2 \rangle = 2(\mathcal{A} - \mathcal{B} + i\mathcal{N})\langle a^2 \rangle. \tag{37}$$

Solving these we can get the results,

$$\langle a \rangle_t = \sqrt{G}\langle a \rangle_0, \tag{38}$$

$$\langle a^\dagger a \rangle_t = G_0 \langle a^\dagger a \rangle_0 + (G_0 - 1)\frac{\mathcal{A}}{\mathcal{A} - \mathcal{B}}, \tag{39}$$

$$\langle a^2 \rangle_t = G\langle a^2 \rangle_0, \tag{40}$$

where $G_0 = \exp[2(\mathcal{A} - \mathcal{B})t], G = G_0 \exp[2i\mathcal{N}t]$. Then it can be shown that

$$\langle X_\theta \rangle_t = \sqrt{G}\langle X_\theta \rangle_0, \tag{41}$$

$$\langle \Delta X_\theta^2 \rangle_t = G_0 \langle \Delta X_{\theta - \mathcal{N}t}^2 \rangle_0 + \frac{1}{4}(G_0 - 1)\frac{\mathcal{A} + \mathcal{B}}{\mathcal{A} - \mathcal{B}}. \tag{42}$$

Here we use the symbol $\langle \Delta X_{\theta - \mathcal{N}t}^2 \rangle_0$ to denote

$$\langle \Delta X_{\theta - \mathcal{N}t}^2 \rangle_0 = \frac{1}{4}\left\{\left(\langle a^{\dagger 2} \rangle_0 - \langle a^\dagger \rangle_0^2\right)e^{2i\theta}e^{-2i\mathcal{N}t} + \left(\langle a^2 \rangle_0 - \langle a \rangle_0^2\right)e^{-2i\theta}e^{2i\mathcal{N}t} + \langle a^\dagger a \rangle_0 + \langle aa^\dagger \rangle_0 - 2\langle a^\dagger \rangle_0 \langle a \rangle_0\right\}. \tag{43}$$

Note that in addition to the gain or loss coefficient $(\mathcal{A} - \mathcal{B})t$, the quadratures also undergo a phase shift $\mathcal{N}t$. As in the resonant case, this result differs significantly from that of the SC-CM. Similarly, only in the limiting cases of pure amplification or attenuation, is it possible to construct an SC-CM model with the exception that the amplitude gain or attenuation factor contains a phase-shift coefficient.

## IV. FOUR-LEVEL ATOMIC SYSTEM

Consider next the interaction of a field with frequency $\upsilon$ and a four-level atomic system as shown schematically in Fig. 6. Briefly, it consists of two transitions, one of which would produce a broad gain spectrum, whereas the other would yield a narrow dip in the gain spectrum. As such, such a system can function as an NDM.

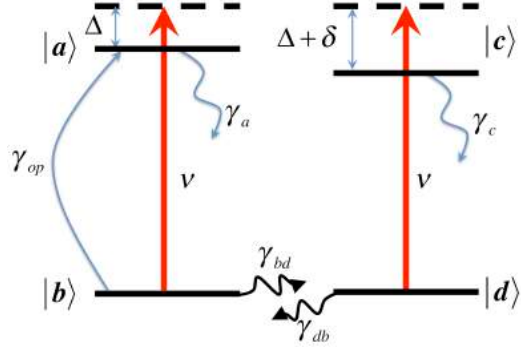

Fig. 6 Four-level atomic system

The Hamiltonian for the atom-field system in the interaction picture after addition of complex terms is as follows:

$$\tilde{\mathcal{H}}' = \hbar\left(-\Delta_a - \frac{i}{2}\gamma_a\right)|a\rangle\langle a| - \frac{i}{2}\hbar\left(\gamma_{op} + \gamma_{bd}\right)|b\rangle\langle b| + \left(-\Delta_c - \frac{i}{2}\gamma_c\right)|c\rangle\langle c| - \frac{i}{2}\hbar\gamma_{db}|d\rangle\langle d| \quad (44)$$
$$+ \hbar g(\sigma_{ab}a + a^{\dagger}\sigma_{ba} + \sigma_{cd}a + a^{\dagger}\sigma_{dc}).$$

where $\Delta_a = \upsilon - (\omega_a - \omega_b) \equiv \Delta$, $\Delta_c = \upsilon - (\omega_c - \omega_d) \equiv \Delta + \delta$. The matrix elements are

$$\tilde{\mathcal{H}}'_{an,an} = -\hbar(\Delta_a + i\gamma_a/2),\ \tilde{\mathcal{H}}'_{bn+1,bn+1} = -i\hbar\left(\gamma_{op} + \gamma_{bd}\right)/2,\ \tilde{\mathcal{H}}'_{cn,cn} = -\hbar(\Delta_c + i\gamma_c/2),\ \tilde{\mathcal{H}}'_{dn+1,dn+1} = -i\hbar\gamma_{db}/2, \quad (45)$$

$$\tilde{\mathcal{H}}'_{an,bn+1} = \hbar g\sqrt{n+1} = \tilde{\mathcal{H}}'_{bn+1,an} = \tilde{\mathcal{H}}'^{*}_{bn+1,an},\ \tilde{\mathcal{H}}'_{cn,dn+1} = \hbar g\sqrt{n+1} = \tilde{\mathcal{H}}'_{dn+1,cn} = \tilde{\mathcal{H}}'^{*}_{dn+1,cn}. \quad (46)$$

The equations of motion for the density matrix are then derived from Eq. (2),

$$\dot{\rho}_{an,an'} = \gamma_{op}\rho_{bn,bn'} - \gamma_a\rho_{an,an'} - \frac{i}{\hbar}(\tilde{\mathcal{H}}'_{an,bn+1}\rho_{bn+1,an'} - \rho_{an,bn'+1}\tilde{\mathcal{H}}'^{*}_{bn'+1,an'}), \quad (47a)$$

$$\dot{\rho}_{an,bn'+1} = -(\gamma_{ab} - i\Delta)\rho_{an,bn'+1} - \frac{i}{\hbar}(\tilde{\mathcal{H}}'_{an,bn+1}\rho_{bn+1,bn'+1} - \rho_{an,an'}\tilde{\mathcal{H}}'^{*}_{an',bn'+1}), \quad (47b)$$

$$\dot{\rho}_{an,cn'} = -(\gamma_{ac} + i\delta)\rho_{an,cn'} - \frac{i}{\hbar}(\tilde{\mathcal{H}}'_{an,bn+1}\rho_{bn+1,cn'} - \rho_{an,dn'+1}\tilde{\mathcal{H}}'^{*}_{dn'+1,cn'}), \quad (47c)$$

$$\dot{\rho}_{an,dn'+1} = -(\gamma_{ad} - i\Delta)\rho_{an,dn'+1} - \frac{i}{\hbar}(\tilde{\mathcal{H}}'_{an,bn+1}\rho_{bn+1,dn'+1} - \rho_{an,cn'}\tilde{\mathcal{H}}'^{*}_{cn',dn'+1}), \quad (47d)$$

$$\dot{\rho}_{bn+1,an'} = -(\gamma_{ab} + i\Delta)\rho_{bn+1,an'} - \frac{i}{\hbar}(\tilde{\mathcal{H}}'_{bn+1,an}\rho_{an,an'} - \rho_{bn+1,bn'+1}\tilde{\mathcal{H}}'^{*}_{bn'+1,an'}), \tag{47e}$$

$$\dot{\rho}_{bn+1,bn'+1} = \gamma_a \rho_{an+1,an'+1} - (\gamma_{op} + \gamma_{bd})\rho_{bn+1,bn'+1} + \gamma_{db}\rho_{dn+1,dn'+1} - \frac{i}{\hbar}(\tilde{\mathcal{H}}'_{bn+1,an}\rho_{an,bn'+1} - \rho_{bn+1,an'}\tilde{\mathcal{H}}'^{*}_{an',bn'+1}), \tag{47f}$$

$$\dot{\rho}_{bn+1,cn'} = -[\gamma_{bc} + i(\Delta+\delta)]\rho_{bn+1,cn'} - \frac{i}{\hbar}(\tilde{\mathcal{H}}'_{bn+1,an}\rho_{an,cn'} - \rho_{bn+1,dn'+1}\tilde{\mathcal{H}}'^{*}_{dn'+1,cn'}), \tag{47g}$$

$$\dot{\rho}_{bn+1,dn'+1} = -\gamma_{bd}\rho_{bn+1,dn'+1} - \frac{i}{\hbar}(\tilde{\mathcal{H}}'_{bn+1,an}\rho_{an,dn'+1} - \rho_{bn+1,cn'}\tilde{\mathcal{H}}'^{*}_{cn',dn'+1}), \tag{47h}$$

$$\dot{\rho}_{cn,an'} = -(\gamma_{ac} - i\delta)\rho_{cn,an'} - \frac{i}{\hbar}(\tilde{\mathcal{H}}'_{cn,dn+1}\rho_{dn+1,an'} - \rho_{cn,bn'+1}\tilde{\mathcal{H}}'^{*}_{bn'+1,an'}), \tag{47i}$$

$$\dot{\rho}_{cn,bn'+1} = -[\gamma_{bc} - i(\Delta+\delta)]\rho_{cn,bn'+1} - \frac{i}{\hbar}(\tilde{\mathcal{H}}'_{cn,dn+1}\rho_{dn+1,bn'+1} - \rho_{cn,an'}\tilde{\mathcal{H}}'^{*}_{an',bn'+1}), \tag{47j}$$

$$\dot{\rho}_{cn,cn'} = -\gamma_c \rho_{cn,cn'} - \frac{i}{\hbar}(\tilde{\mathcal{H}}'_{cn,dn+1}\rho_{dn+1,cn'} - \rho_{cn,dn'+1}\tilde{\mathcal{H}}'^{*}_{dn'+1,cn'}), \tag{47k}$$

$$\dot{\rho}_{cn,dn'+1} = -[\gamma_{cd} - i(\Delta+\delta)]\rho_{cn,dn'+1} - \frac{i}{\hbar}(\tilde{\mathcal{H}}'_{cn,dn+1}\rho_{dn+1,dn'+1} - \rho_{cn,cn'}\tilde{\mathcal{H}}'^{*}_{cn',dn'+1}), \tag{47l}$$

$$\dot{\rho}_{dn+1,an'} = -(\gamma_{ad} + i\Delta)\rho_{dn+1,an'} - \frac{i}{\hbar}(\tilde{\mathcal{H}}'_{dn+1,cn}\rho_{cn,an'} - \rho_{dn+1,bn'+1}\tilde{\mathcal{H}}'^{*}_{bn'+1,an'}), \tag{47m}$$

$$\dot{\rho}_{dn+1,bn'+1} = -\gamma_{bd}\rho_{dn+1,bn'+1} - \frac{i}{\hbar}(\tilde{\mathcal{H}}'_{dn+1,cn}\rho_{cn,bn'+1} - \rho_{dn+1,an'}\tilde{\mathcal{H}}'^{*}_{an',bn'+1}), \tag{47n}$$

$$\dot{\rho}_{dn+1,cn'} = -[\gamma_{cd} + i(\Delta+\delta)]\rho_{dn+1,cn'} - \frac{i}{\hbar}(\tilde{\mathcal{H}}'_{dn+1,cn}\rho_{cn,cn'} - \rho_{dn+1,dn'+1}\tilde{\mathcal{H}}'^{*}_{dn'+1,cn'}), \tag{47o}$$

$$\dot{\rho}_{dn+1,dn'+1} = \gamma_c \rho_{cn+1,cn'+1} - (\gamma_d + \gamma_{db})\rho_{dn+1,dn'+1} + \gamma_{bd}\rho_{bn+1,bn'+1} - \frac{i}{\hbar}(\tilde{\mathcal{H}}'_{dn+1,cn}\rho_{cn,dn'+1} - \rho_{dn+1,cn'}\tilde{\mathcal{H}}'^{*}_{cn',dn'+1}), \tag{47p}$$

where $\gamma_{\alpha\beta} = (\gamma'_{\alpha} + \gamma'_{\beta})/2, \alpha,\beta = a,b,c,d$ with $\gamma'_a = \gamma_a$, $\gamma'_b = \gamma_{op} + \gamma_{bd}$, $\gamma'_c = \gamma_c$, and $\gamma'_d = \gamma_{db}$. It can be seen that Eqs. (47a), (47f), and (47p) contain terms $\rho_{bn,bn'}$, $\rho_{an+1,an'+1}$ and $\rho_{cn+1,cn'+1}$, which fall outside of the *n* manifold of density matrix elements (following the indexing notation we introduced earlier). Note that one of these ($\rho_{bn,bn'}$) belongs to the (*n* − 1) manifold, whereas the other two ($\rho_{an+1,an'+1}$ and $\rho_{cn+1,cn'+1}$) both belong to the (*n* + 1) manifold. Using the constraint that $\rho_{an,an'} + \rho_{bn,bn'} + \rho_{cn,cn'} + \rho_{dn,dn'} = \tilde{\rho}_{nn'}$ cannot totally remove the coupling between different manifolds, unlike the case in Sec. III. We make use of the steady state solutions of the atomic system (when the probe field is zero), such as

$$\rho_{aa}^{(0)} = \frac{\gamma_{op}\gamma_{db}}{\gamma_{op}\gamma_{db} + \gamma_a(\gamma_{bd} + \gamma_{db})}, \rho_{bb}^{(0)} = \frac{\gamma_a\gamma_{db}}{\gamma_{op}\gamma_{db} + \gamma_a(\gamma_{bd} + \gamma_{db})}, \rho_{cc}^{(0)} = 0, \rho_{dd}^{(0)} = \frac{\gamma_a\gamma_{b0}}{\gamma_{op}\gamma_{db} + \gamma_a(\gamma_{bd} + \gamma_{db})}, \quad (48)$$

to make the approximation that

$$\rho_{bn,bn'} = \tilde{\rho}_{nn'} - \rho_{an,an'} - \rho_{cn,cn'} - \rho_{dn,dn'} \simeq \tilde{\rho}_{nn'} - \rho_{an,an'} - \rho_{cc}^{(0)}\tilde{\rho}_{nn'} - \rho_{dd}^{(0)}\tilde{\rho}_{nn'} = \left(1 - \rho_{cc}^{(0)} - \rho_{dd}^{(0)}\right)\tilde{\rho}_{nn'} - \rho_{an,an'}, \quad (49)$$

which is valid for very weak fields. Using the same argument, we can approximate that

$$\rho_{an+1,an'+1} \simeq \left(1 - \rho_{cc}^{(0)} - \rho_{dd}^{(0)}\right)\tilde{\rho}_{n+1,n'+1} - \rho_{bn+1,bn'+1}, \quad (50)$$

$$\rho_{cn+1,cn'+1} \simeq \left(1 - \rho_{aa}^{(0)} - \rho_{bb}^{(0)}\right)\tilde{\rho}_{n+1,n'+1} - \rho_{dn+1,dn'+1}. \quad (51)$$

The use of these approximations is validated in Appendix B 2 by showing that the susceptibilities calculated using the ME model with these approximations agree with the semiclassical results. Solving the set of equations with the same method as used in Eqs. (A16)–(A19) in Appendix A, we get the matrix elements in the linear regime as

$$\rho_{an,bn'+1} = \frac{ig\gamma_{db}\left(\sqrt{n'+1}\gamma_{op}\tilde{\rho}_{nn'} - \sqrt{n+1}\gamma_a\tilde{\rho}_{n+1,n'+1}\right)}{(\gamma_{ab} + i\Delta)\left[\gamma_a(\gamma_{bd} + \gamma_{db}) + \gamma_{op}\gamma_{db}\right]}, \; \rho_{bn+1,an'} = \frac{-ig\gamma_{db}\left(\sqrt{n+1}\gamma_{op}\tilde{\rho}_{nn'} - \sqrt{n'+1}\gamma_a\tilde{\rho}_{n+1,n'+1}\right)}{(\gamma_{ab} - i\Delta)\left[\gamma_a(\gamma_{bd} + \gamma_{db}) + \gamma_{op}\gamma_{db}\right]}, \quad (52)$$

$$\rho_{cn,dn'+1} = -\frac{ig\sqrt{n+1}\gamma_a\gamma_{bd}\tilde{\rho}_{n+1,n'+1}}{\left[\gamma_{cd} + i(\Delta + \delta)\right]\left[\gamma_a(\gamma_{bd} + \gamma_{db}) + \gamma_{op}\gamma_{db}\right]}, \; \rho_{dn+1,cn'} = \frac{ig\sqrt{n'+1}\gamma_a\gamma_{bd}\tilde{\rho}_{n+1,n'+1}}{\left[\gamma_{cd} - i(\Delta + \delta)\right]\left[\gamma_a(\gamma_{bd} + \gamma_{db}) + \gamma_{op}\gamma_{db}\right]}. \quad (53)$$

Tracing over the atomic states in Eq. (2), we get

$$\dot{\tilde{\rho}}_{nn'} = -\frac{i}{\hbar}\begin{pmatrix}\tilde{\mathcal{H}}'_{an,bn+1}\rho_{bn+1,an'} - \rho_{an,bn'+1}\tilde{\mathcal{H}}'^{*}_{bn'+1,an'} + \tilde{\mathcal{H}}'_{bn,an-1}\rho_{an-1,bn'} - \rho_{bn,an'-1}\tilde{\mathcal{H}}'^{*}_{an'-1,bn'} \\ +\tilde{\mathcal{H}}'_{cn,dn+1}\rho_{dn+1,cn'} - \rho_{cn,dn'+1}\tilde{\mathcal{H}}'^{*}_{dn'+1,cn'} + \tilde{\mathcal{H}}'_{dn,cn-1}\rho_{cn-1,dn'} - \rho_{dn,cn'-1}\tilde{\mathcal{H}}'^{*}_{cn'-1,dn'}\end{pmatrix}, \quad (54)$$

which is essentially the same as

$$\dot{\tilde{\rho}}_{nn'} = -\frac{i}{\hbar}\begin{pmatrix}\tilde{\mathcal{H}}_{an,bn+1}\rho_{bn+1,an'} - \rho_{an,bn'+1}\tilde{\mathcal{H}}_{bn'+1,an'} + \tilde{\mathcal{H}}_{bn,an-1}\rho_{an-1,bn'} - \rho_{bn,an'-1}\tilde{\mathcal{H}}_{an'-1,bn'} \\ +\tilde{\mathcal{H}}_{cn,dn+1}\rho_{dn+1,cn'} - \rho_{cn,dn'+1}\tilde{\mathcal{H}}_{dn'+1,cn'} + \tilde{\mathcal{H}}_{dn,cn-1}\rho_{cn-1,dn'} - \rho_{dn,cn'-1}\tilde{\mathcal{H}}_{cn'-1,dn'}\end{pmatrix}. \quad (55)$$

Plugging in the solutions in Eqs. (52) and (53), we derive that

$$\dot{\tilde{\rho}}_{nn'} = -\mathcal{A}\left[(n+n'+2)\tilde{\rho}_{nn'} - 2\sqrt{nn'}\tilde{\rho}_{n-1,n'-1}\right] - (\mathcal{B} + \mathcal{D})\left[(n+n')\tilde{\rho}_{nn'} - 2\sqrt{(n+1)(n'+1)}\tilde{\rho}_{n+1,n'+1}\right] \\ + i\mathcal{N}(n-n')\tilde{\rho}_{nn'}, \quad (56)$$

where

$$\mathcal{A} = \frac{g^2\gamma_{ab}\gamma_{op}\gamma_{db}}{\left(\gamma_{ab}^2 + \Delta^2\right)\left[\gamma_a(\gamma_{bd} + \gamma_{db}) + \gamma_{op}\gamma_{db}\right]}, \quad \mathcal{B} = \frac{g^2\gamma_{ab}\gamma_a\gamma_{db}}{\left(\gamma_{ab}^2 + \Delta^2\right)\left[\gamma_a(\gamma_{bd} + \gamma_{db}) + \gamma_{op}\gamma_{db}\right]}, \quad (57)$$

$$\mathcal{D} = \frac{g^2 \gamma_{cd} \gamma_a \gamma_{bd}}{\left[\gamma_{cd}^2 + (\Delta+\delta)^2\right]\left[\gamma_a(\gamma_{bd}+\gamma_{db})+\gamma_{op}\gamma_{db}\right]}, \tag{58}$$

$$\mathcal{N} = \frac{g^2}{\gamma_{ab}^2 + \Delta^2} \frac{(\gamma_{op}-\gamma_a)\gamma_{db}}{\gamma_a(\gamma_{bd}+\gamma_{db})+\gamma_{op}\gamma_{db}} \Delta - \frac{g^2}{\gamma_{cd}^2+(\Delta+\delta)^2} \frac{\gamma_a \gamma_{bd}}{\gamma_a(\gamma_{bd}+\gamma_{db})+\gamma_{op}\gamma_{db}}(\Delta+\delta). \tag{59}$$

Alternatively, as shown in Sec. III.A, Eq. (56) can be written as

$$\dot{\tilde{\rho}} = -\mathcal{A}(aa^\dagger\tilde{\rho} - 2a^\dagger\tilde{\rho}a + \tilde{\rho}aa^\dagger) - (\mathcal{B}+\mathcal{D})(a^\dagger a\tilde{\rho} - 2a\tilde{\rho}a^\dagger + \tilde{\rho}a^\dagger a) + i\mathcal{N}(a^\dagger a\tilde{\rho} - \tilde{\rho}a^\dagger a). \tag{60}$$

We can then derive the equations of motion for $a^{\dagger m} a^n$,

$$\frac{d}{dt}\langle a \rangle = (\mathcal{A} - \mathcal{B} - \mathcal{D} + i\mathcal{N})\langle a \rangle, \tag{61}$$

$$\frac{d}{dt}\langle a^\dagger a \rangle = 2(\mathcal{A} - \mathcal{B} - \mathcal{D})\langle a \rangle + 2\mathcal{A}, \tag{62}$$

$$\frac{d}{dt}\langle a^2 \rangle = 2(\mathcal{A} - \mathcal{B} - \mathcal{D} + i\mathcal{N})\langle a^2 \rangle, \tag{63}$$

whose solutions are

$$\langle a \rangle_t = \sqrt{G}\langle a \rangle_0 \tag{64}$$

$$\langle a^\dagger a \rangle_t = G_0 \langle a^\dagger a \rangle_0 + (G_0 - 1)\frac{\mathcal{A}}{\mathcal{A} - \mathcal{B} - \mathcal{D}} \tag{65}$$

$$\langle a^2 \rangle_t = G\langle a^2 \rangle_0 \tag{66}$$

where $G_0 = \exp[2(\mathcal{A}-\mathcal{B}-\mathcal{D})t]$, $G = G_0 \exp[2i\mathcal{N}t]$. As a result,

$$\langle X_\theta \rangle_t = \sqrt{G}\langle X_\theta \rangle_0, \tag{67}$$

$$\langle \Delta X_\theta^2 \rangle_t = G_0 \langle \Delta X_{\theta-\mathcal{N}t}^2 \rangle_0 + (G_0 - 1)\frac{\mathcal{A}+\mathcal{B}+\mathcal{D}}{4(\mathcal{A}-\mathcal{B}-\mathcal{D})}. \tag{68}$$

In Fig. 7, we plot $\mathcal{A} - \mathcal{B} - \mathcal{D} \equiv \varepsilon$ and $\mathcal{N}$, which is proportional to the gain and phase shift, respectively, as a function of $\Delta$. Figure 7(a) shows a dip at $\Delta = -\delta$ in the broad gain profile, and it can be seen from Fig. 7(b) that the system entails a negative dispersion around $\Delta = -\delta$. The additional QN term in Eq. (68) differs from that in Eq. (26) predicted by the SC-CM. Therefore, the SC-CM does not apply to the four-level atomic system in describing the QN. This disagreement is catastrophic since the SC-CM prediction can in no way be thought of as an approximation of the actual result.

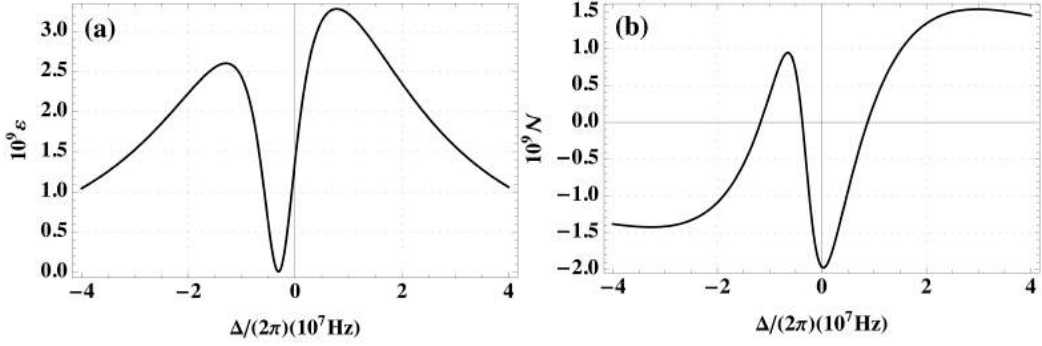

Fig. 7 Plot of $\varepsilon$ and $\mathcal{N}$ as a function $\Delta/(2\pi)$ in the four-level system in Fig. 6.

Let us consider the noise at the dip ($\Delta = -\delta$). Denote $T$ as the interaction time through the medium and the corresponding gain as $G_T = \exp(2\varepsilon T)$. At $\Delta = -\delta$, there is perfect transparency ($G_T = 1$) so that $\varepsilon = 0$. However, for the additional noise term in the variance as seen in Eq. (68), although $(G_T - 1)$ goes to zero, the denominator $\varepsilon$ also goes to zero. Therefore the total noise is not zero as predicted by the SC-CM but a finite and large number, which can be determined by considering the limit $\varepsilon \to 0$,

$$\lim_{\varepsilon \to 0}(G_T - 1)\frac{\mathcal{A}+\mathcal{B}+\mathcal{D}}{4(\mathcal{A}-\mathcal{B}-\mathcal{D})} = \lim_{\varepsilon \to 0}\frac{e^{2\varepsilon T}-1}{\varepsilon}\frac{1}{4}(\mathcal{A}+\mathcal{B}+\mathcal{D}) = \frac{1}{2}(\mathcal{A}+\mathcal{B}+\mathcal{D})T. \tag{69}$$

As a result, this four-level atomic system is not suitable for our WLC-SR scheme [15], which requires very low QN. In Sec. VI, we will describe a five-level atomic system that has a negative dispersion suitable for use in the WLC-SR scheme and much lower QN than that of the four-level system.

## V. EIT

In the preceding section, we considered a system where the QN is found to be nonvanishing even when the mean net gain or absorption is zero. However, this is not necessarily true for all systems. As an example of an exception, we consider next a system where the SC-CM and the ME model agree, and thus the QN vanishes when the mean absorption is zero. It is also our inspiration for the GEIT system in Sec. VI. This system is illustrated schematically in Fig. 8. This is known as the Λ-type EIT system where a probe field excites atoms from level $|b\rangle$ to level $|a\rangle$ and a coherent pump field excites atoms from $|c\rangle$ to $|a\rangle$. We treat the probe field quantum mechanically. Assume that the pump field is at resonance with the $|a\rangle - |c\rangle$ transition and that the decay rates from level $|a\rangle$ to levels $|b\rangle$ and $|c\rangle$ are the same, i.e., $\Gamma_{ab} = \Gamma_{ac} = \gamma/2$. Then the Hamiltonian after the transformation to the interaction picture and addition of complex terms can be written as:

$$\tilde{\mathcal{H}}' = -\frac{i}{2}\hbar\gamma|a\rangle\langle a| + \hbar\Delta|b\rangle\langle b| + \hbar g(|a\rangle\langle b|a + a^\dagger|b\rangle\langle a|) - \frac{1}{2}\hbar\Omega_p(|a\rangle\langle c| + |c\rangle\langle a|). \tag{70}$$

Here $\Delta = \upsilon - (\omega_a - \omega_b)$ and $\Omega_p$ is the Rabi frequency of the pump field.

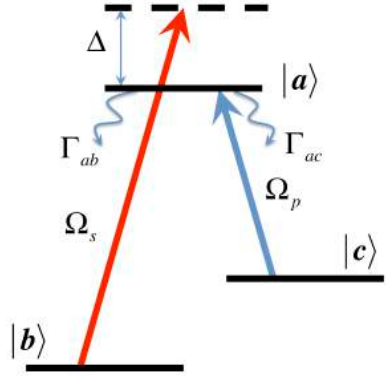

Fig. 8 Three-level, Λ-type EIT configuration.

It can be shown that $\tilde{\mathcal{H}}'_{an,an} = -i\hbar\gamma/2$, $\tilde{\mathcal{H}}'_{bn+1,bn+1} = -\hbar\Delta$, $\tilde{\mathcal{H}}'_{an,bn+1} = \hbar g\sqrt{n+1} = \tilde{\mathcal{H}}'_{bn+1,an} = \tilde{\mathcal{H}}'^{*}_{bn+1,an}$, and $\tilde{\mathcal{H}}'_{an,cn} = -\hbar\Omega_p e^{-i\phi} = \tilde{\mathcal{H}}'^{*}_{cn,an}$, whereas the other elements are zero. The equations of motion for the density matrix elements can be derived from Eq. (2),

$$\dot\rho_{an,an'} = -\gamma\rho_{an,an'} - \frac{i}{\hbar}(\tilde{\mathcal{H}}'_{an,bn+1}\rho_{bn+1,an'} - \rho_{an,bn'+1}\tilde{\mathcal{H}}'^{*}_{bn'+1,an'} + \tilde{\mathcal{H}}'_{an,cn}\rho_{cn,an'} - \rho_{an,cn'}\tilde{\mathcal{H}}'^{*}_{cn',an'}), \tag{71}$$

$$\dot\rho_{an,bn'+1} = -\left(\frac{\gamma}{2} - i\Delta\right)\rho_{an,bn'+1} - \frac{i}{\hbar}(\tilde{\mathcal{H}}'_{an,bn+1}\rho_{bn+1,bn'+1} - \rho_{an,an'}\tilde{\mathcal{H}}'^{*}_{an',bn'+1} + \tilde{\mathcal{H}}'_{an,cn}\rho_{cn,bn'+1}), \tag{72}$$

$$\dot\rho_{an,cn'} = -\frac{\gamma}{2}\rho_{an,bn'+1} - \frac{i}{\hbar}(\tilde{\mathcal{H}}'_{an,bn+1}\rho_{bn+1,cn'} - \rho_{an,an'}\tilde{\mathcal{H}}'^{*}_{an',cn'} + \tilde{\mathcal{H}}'_{an,cn}\rho_{cn,cn'}), \tag{73}$$

$$\dot\rho_{bn+1,an'} = -\left(\frac{\gamma}{2} + i\Delta\right)\rho_{bn+1,an'} - \frac{i}{\hbar}(\tilde{\mathcal{H}}'_{bn+1,an}\rho_{an,an'} - \rho_{bn+1,bn'+1}\tilde{\mathcal{H}}'^{*}_{bn'+1,an'} - \rho_{bn+1,cn'}\tilde{\mathcal{H}}'^{*}_{cn',an'}), \tag{74}$$

$$\dot\rho_{bn+1,bn'+1} = \frac{\gamma}{2}\rho_{an+1,an'+1} - \frac{i}{\hbar}(\tilde{\mathcal{H}}'_{bn+1,an}\rho_{an,bn'+1} - \rho_{bn+1,an'}\tilde{\mathcal{H}}'^{*}_{an',bn'+1}), \tag{75}$$

$$\dot{\rho}_{bn+1,cn'} = -i\Delta\rho_{bn+1,cn'} - \frac{i}{\hbar}(\tilde{\mathcal{H}}'_{bn+1,an}\rho_{an,cn'} - \rho_{bn+1,an'}\tilde{\mathcal{H}}'^*_{an',cn'}), \tag{76}$$

$$\dot{\rho}_{cn,an'} = -\frac{\gamma}{2}\rho_{cn,an'} - \frac{i}{\hbar}(\tilde{\mathcal{H}}'_{cn,an}\rho_{an,an'} - \rho_{cn,cn'}\tilde{\mathcal{H}}'^*_{cn',an'} - \rho_{cn,bn'+1}\tilde{\mathcal{H}}'^*_{bn'+1,an'}), \tag{77}$$

$$\dot{\rho}_{cn,bn'+1} = i\Delta\rho_{cn,bn'+1} - \frac{i}{\hbar}(\tilde{\mathcal{H}}'_{cn,an}\rho_{an,bn'+1} - \rho_{cn,an'}\tilde{\mathcal{H}}'^*_{an',bn'+1}), \tag{78}$$

$$\dot{\rho}_{cn,cn'} = \frac{\gamma}{2}\rho_{an,an'} - \frac{i}{\hbar}(\tilde{\mathcal{H}}'_{cn,an}\rho_{an,cn'} - \rho_{cn,an'}\tilde{\mathcal{H}}'^*_{an',cn'}), \tag{79}$$

Just as in the case considered in the previous section, we again see that the relation $\rho_{an,an'} + \rho_{bn,bn'} + \rho_{cn,cn'} = \tilde{\rho}_{n,n'}$ is not enough to decouple the neighboring manifolds. As such, we use the same type of approximation used earlier in deriving Eqs. (49)–(51). Specifically, we use the steady-state solution when the probe field is absent,

$$\rho^{(0)}_{aa} = 0, \rho^{(0)}_{bb} = 1, \rho^{(0)}_{cc} = 0. \tag{80}$$

The validity of the approximations is shown in Appendix B 3. Using this result we can now write

$$\rho_{an+1,an'+1} = \tilde{\rho}_{n+1,n'+1} - \rho_{bn+1,bn'+1} - \rho_{cn+1,cn'+1} \simeq \tilde{\rho}_{n+1,n'+1} - \rho_{bn+1,bn'+1}. \tag{81}$$

Then we can solve the set of equations within each manifold, and plug the result into

$$\dot{\tilde{\rho}}_{nn'} = -\frac{i}{\hbar}\left(\tilde{\mathcal{H}}'_{an,bn+1}\rho_{bn+1,an'} - \rho_{an,bn'+1}\tilde{\mathcal{H}}'^*_{bn'+1,an'} + \tilde{\mathcal{H}}'_{bn,an-1}\rho_{an-1,bn'} - \rho_{bn,an'-1}\tilde{\mathcal{H}}'^*_{an'-1,bn'}\right). \tag{82}$$

This yields

$$\dot{\tilde{\rho}}_{nn'} = -\mathcal{B}\left[(n+n')\tilde{\rho}_{nn'} - 2\sqrt{(n+1)(n'+1)}\tilde{\rho}_{n+1,n'+1}\right] + i\mathcal{N}(n-n')\tilde{\rho}_{nn'}, \tag{83}$$

where

$$\mathcal{B} = \frac{g^2\Delta^2\gamma/2}{\gamma^2\Delta^2/4 + (\Delta^2 - \Omega_p^2/4)^2}, \quad \mathcal{N} = -\frac{g^2\Delta(\Delta^2 - \Omega_p^2/4)}{\gamma^2\Delta^2/4 + (\Delta^2 - \Omega_p^2/4)^2}, \tag{84}$$

These quantities are plotted as functions of $\Delta$ in Fig. 9. Equation (83) can also be written in the form of

$$\dot{\tilde{\rho}} = -\mathcal{B}(a^\dagger a\tilde{\rho} - 2a\tilde{\rho}a^\dagger + \tilde{\rho}a^\dagger a) + i\mathcal{N}\left(a^\dagger a\tilde{\rho} - \tilde{\rho}a^\dagger a\right). \tag{85}$$

The evolution of moments of $a$ and $a^\dagger$ are determined by

$$\frac{d}{dt}\langle a\rangle = -(\mathcal{B} - i\mathcal{N})\langle a\rangle, \tag{86}$$

$$\frac{d}{dt}\langle a^\dagger a\rangle = -2\mathcal{B}\langle a^\dagger a\rangle, \tag{87}$$

$$\frac{d}{dt}\langle a^2\rangle = -2(\mathcal{B}-i\mathcal{N})\langle a^2\rangle, \tag{88}$$

the solutions of which are as follows:

$$\langle a\rangle_t = \sqrt{G}\langle a\rangle_0, \tag{89}$$

$$\langle a^\dagger a\rangle_t = G_0\langle a^\dagger a\rangle_0, \tag{90}$$

$$\langle a^2\rangle_t = G\langle a^2\rangle_0, \tag{91}$$

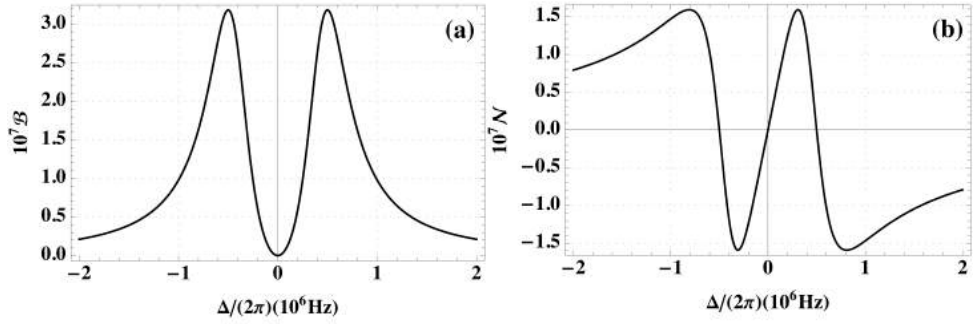

Fig. 9. Plot of (a) $\mathcal{B}$ and (b) $\mathcal{N}$ as a function of $\Delta/(2\pi)$ in the case of $\Lambda$-type EIT scheme.

where $G_0 = \exp[-2\mathcal{B}t], G = G_0\exp[2i\mathcal{N}t]$. As a result,

$$\langle X_\theta\rangle_t = \sqrt{G}\langle X_\theta\rangle_0 \tag{92}$$

$$\langle \Delta X_\theta^2\rangle_t = G_0\langle \Delta X_{\theta-\mathcal{N}t}^2\rangle_0 + \frac{1}{4}(1-G_0), \tag{93}$$

As shown in Eqs. (92) and (93), the results of the ME in the EIT system agree with those from the SC-CM with an additional phase shift. At zero detuning ($\Delta = 0$), we have both $\mathcal{B} = 0$ and $\mathcal{N} = 0$, so that $G_0 = 1$ and the additional noise in the variance of the quadrature is zero. We expect that in steady state the system is in the dark state with no excitation in the level $|a\rangle$ at zero detuning:

$$|D\rangle = \frac{1}{\sqrt{\Omega_p^2+\Omega_s^2}}(\Omega_p|b\rangle - \Omega_s|c\rangle), \tag{94}$$

where $\Omega_p$ is the Rabi frequency of the pump field, and $\Omega_s$ is the Rabi frequency of the probe field. Since this is a pure state, the expected value of the elements of the density matrix can be written down by inspection:

$$\rho_{aa} = 0, \; \rho_{ab} = \rho_{ab}^* = 0, \; \rho_{ac} = \rho_{ca}^* = 0, \tag{95}$$

$$\rho_{bb} = \frac{\Omega_p^2}{\Omega_p^2 + \Omega_s^2}, \; \rho_{cc} = \frac{\Omega_s^2}{\Omega_p^2 + \Omega_s^2}, \; \rho_{bc} = \rho_{cb} = -\frac{\Omega_p \Omega_s}{\Omega_p^2 + \Omega_s^2}. \tag{96}$$

From the solution of the semiclassical equation of motion for the density matrix computed as Eq. (B22) −(B27) in Appendix 2, the above relations are validated for zero detuning. Therefore the system is indeed in the dark state as in Eq. (94). This explains why the additional noise is zero at zero detuning as shown in Eq. (93), since there are no atoms in the intermediate state $|c\rangle$.

## VI. FIVE-LEVEL GEIT SYSTEM

In Sec. IV, we showed that the NDM realized by a four-level atomic system combining a gain profile with an absorption dip is not suitable for the WLC-SR since the QN is significant even when the net gain is zero. On the other hand, we showed in Sec. V that an EIT system can produce a condition where the QN is zero while the absorption is also zero. However, the dispersion at this condition is positive, thus making the EIT system unsuitable for the WLC-SR scheme. Here, we propose a system that produces an EIT dip superimposed on a broad gain profile. At the center of the dip, the noise of the system is very small, whereas the dispersion is negative. We choose to call this a GEIT system.

The GEIT system consists of a five-level $M$-type configuration where the transitions $|1\rangle - |4\rangle$, $|2\rangle - |4\rangle$ and $|3\rangle - |5\rangle$ are coupled by pump fields $\Omega_1$, $\Omega_2$, and $\Omega_4$, respectively, whereas the transition $|2\rangle - |5\rangle$ is coupled by probe field $\Omega_3$ as shown schematically in Fig. 10(a). A similar five-level system has been studied in Refs. [24] and [25], although under different conditions. State $|4\rangle$ decays to states $|1\rangle$ and $|2\rangle$ at rates $\Gamma_{41}$ and $\Gamma_{42}$, respectively. Similarly, state $|5\rangle$ decays to states $|2\rangle$ and $|3\rangle$ at rates $\Gamma_{52}$ and $\Gamma_{53}$ respectively. Furthermore, we assume that atoms in state $|2\rangle$ decay rapidly to states $|1\rangle$ and $|3\rangle$ at rates $\Gamma_{21}$ and $\Gamma_{23}$, respectively. In practice, these decay rates can be generated via optical pumping by coupling

$|2\rangle$ to other intermediate states that decay to $|1\rangle$ and $|3\rangle$. These decay rates produce a Raman-type population inversion between states $|1\rangle$ and $|2\rangle$ and between states $|3\rangle$ and $|2\rangle$. As such, $\Omega_3$ will experience Raman gain in the presence of $\Omega_4$. Similarly, $\Omega_2$ will experience Raman gain in the presence of $\Omega_1$. However, when both legs ($|1\rangle-|4\rangle-|2\rangle$ and $|3\rangle-|5\rangle-|2\rangle$) are two photon resonant, the Raman transition amplitude from $|1\rangle$ to $|2\rangle$ can cancel the Raman transition amplitude from $|3\rangle$ to $|2\rangle$ under certain circumstances, for example, when the system is in a dark state consisting of a properly weighted linear superposition of states $|1\rangle$ and $|3\rangle$.

To see the behavior of the system at the center of the dip more transparently, it is instructive to consider a reduced system produced via adiabatic elimination of states $|4\rangle$ and $|5\rangle$ (this approximation is for illustration only, and will not be made in the ME analysis as well as the semiclassical analysis to follow in this section). The system is then reduced to a configuration similar to the $\Lambda$-type EIT system as shown in Fig. 10(b). To be concrete, we define $\delta_a \equiv (\delta_1 + \delta_2)/2$, $\Delta_a \equiv (\delta_1 - \delta_2)$, $\delta_b \equiv (\delta_3 + \delta_4)/2$, and $\Delta_b \equiv (\delta_3 - \delta_4)$. Then, the effective Rabi frequencies for the two legs of the reduced $\Lambda$ system are $\Omega_a \simeq \Omega_1\Omega_2/(2\delta_a)$ and $\Omega_b \simeq \Omega_3\Omega_4/(2\delta_b)$ [26]. We assume that $\Delta_a$ is chosen to balance the differential light shift experienced by level $|1\rangle$ [$\Omega_1^2/(4\delta_1)$] and $|2\rangle$ [$\Omega_2^2/(4\delta_2) + \Omega_3^2/(4\delta_3)$], so that the left leg of the reduced transition is resonant. For the other leg, we define $\Delta = \Delta_b - \Delta_{b0}$, where $\Delta = 0$ corresponds to the condition where the value of $\Delta_b$ balances the differential light shift experienced by level $|3\rangle$ [$\Omega_4^2/(4\delta_4)$] and $|2\rangle$ [$\Omega_2^2/(4\delta_2) + \Omega_3^2/(4\delta_3)$]. Then, $\Delta$ represents a net two photon detuning for the reduced $\Lambda$ system.

Under this approximation, for $\Delta = 0$, the system is in a dark state,

$$|D\rangle = (\Omega_b|3\rangle - \Omega_a|1\rangle)/\sqrt{\Omega_a^2 + \Omega_b^2}. \tag{97}$$

Denote the amplitude of state 2 as $c_2$. Then for a small interval of time $\Delta t$, the net change in the amplitude $\delta c_2(\Delta t)|_{net}$ is determined by:

$$\delta c_2(\Delta t)|_{net} = \delta c_2(\Delta t)|_{channel\ a} + \delta c_2(\Delta t)|_{channel\ b}, \tag{98}$$

where the contribution from the excitation of level $|1\rangle$ and that of level $|2\rangle$, respectively, are

$$\delta c_2(\Delta t)\big|_{\text{channel a}} = i\frac{\Omega_a}{2}\Delta t \cdot c_3, \tag{99}$$

$$\delta c_2(\Delta t)\big|_{\text{channel b}} = i\frac{\Omega_b}{2}\Delta t \cdot c_1. \tag{100}$$

Since $c_3 = \Omega_b / \sqrt{\Omega_a^2 + \Omega_b^2}$ and $c_1 = -\Omega_a / \sqrt{\Omega_a^2 + \Omega_b^2}$, the net result is that $\delta c_2(\Delta t)\big|_{\text{net}} = 0$, which means that there is no transition to level $|2\rangle$ and therefore no gain for probe $\Omega_3$. Since this result holds for a small value of $\Delta t$, it holds for any value of $t$, which can be built up by adding small steps of $\Delta t$. However, in the nonresonant case ($\Delta \neq 0$), this cancellation process is not perfect anymore, allowing for $|3\rangle \to |2\rangle$ excitation. Since the population of atoms in level $|3\rangle$ is larger than that in level $|2\rangle$, we have gain for the probe for $\Delta \neq 0$.

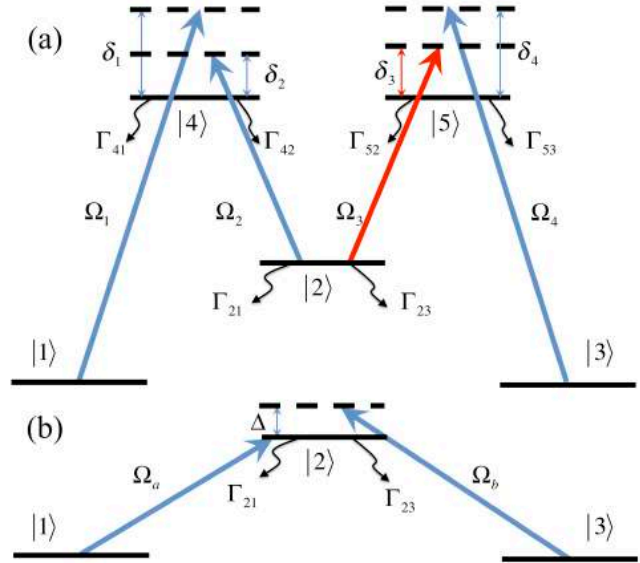

Fig. 10 Schematic illustration of the five-level GEIT system.

We consider first a case where the parameters are $\Omega_1 = \gamma$, $\Omega_2 = 10\gamma$, $\Omega_3 = 10\gamma$, $\Omega_4 = \gamma$, $\delta_1 \simeq \delta_2 \simeq \delta_3 \simeq \delta_4 \simeq 10^3 \gamma$, and $\Gamma_{41} = \Gamma_{42} = \Gamma_{52} = \Gamma_{53} = \Gamma_{21} = \Gamma_{23} = \gamma/2$. Taking into account light shifts when designing the detunings of the fields, the semiclassical result for the complex susceptibility $\chi$ is plotted in Fig. 11, which indeed exhibits a transmission profile with a dip on top of a broader gain and a negative dispersion. This result is obtained by solving the semiclassical density-matrix equation of evolution for the complete five-level system.

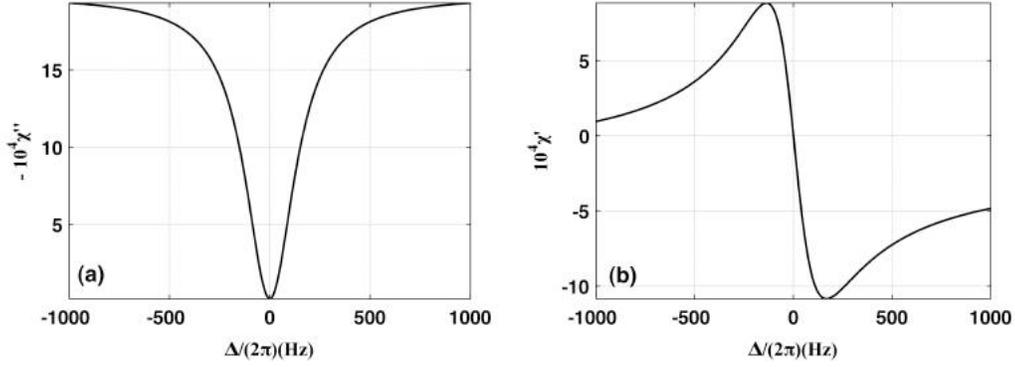

Fig. 11 Plot of (a) imaginary and (b) real part of the complex susceptibility as a function of detuning for the GEIT system from semiclassical calculation. Here $\gamma/(2\pi) = 6$MHz, $\Gamma_{41} = \Gamma_{42} = \Gamma_{52} = \Gamma_{53} = \Gamma_{21} = \Gamma_{23} = \gamma/2$, $\delta_1 \simeq \delta_2 \simeq \delta_3 \simeq \delta_4 \simeq 10^3\gamma$, and $\Omega_1 = \Omega_4 = \gamma$, $\Omega_2 = \Omega_3 = 10\gamma$.

As noted above, the reduction of a three-level system (such as the $|1\rangle - |4\rangle - |2\rangle$ leg of the *M*-system) to a two-level system involves adiabatic elimination of the intermediate state, which is an approximation. As such, we find that the steady state solution for the *M*-system at the center of the dip differs slightly from what is expected for a pure dark state consisting of states $|1\rangle$ and $|3\rangle$ only. For example, the populations of states $|4\rangle$, $|5\rangle$, and $|2\rangle$ are not completely vanishing. However, for the parameters used in producing the plot of Fig. 11, we find that $\rho_{44} \simeq \rho_{55} \simeq 1.25 \times 10^{-7}$ and $\rho_{22} \simeq 3.7 \times 10^{-7}$, which are very close to zero. We also find that $\rho_{13} \simeq -0.4950$ as shown in Fig. 12, which is very close to the value $-0.5$ expected for the dark state.

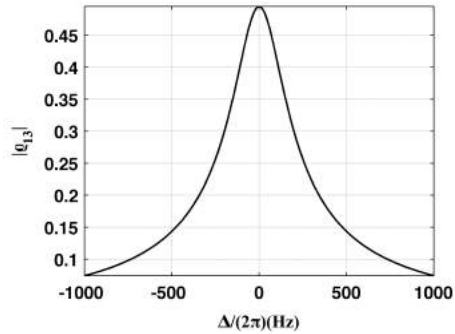

Fig. 12 Plot of $|\rho_{13}|$ versus $\Delta/2\pi$ for the GEIT system. The parameters are the same as those used in Fig. 11.

However, the set of parameters used in this example is not suitable for the WLC-SR. Since the gravitational wave signal is very weak in the WLC-SR, the probe field $\Omega_3$ is vanishingly small. For such a small value of $\Omega_3$, the approximation involved in eliminating states $|4\rangle$ and $|5\rangle$ adiabatically may not necessarily hold, and the steady-state solution at the center of the dip may not necessarily correspond to the dark state given by Eq. (97). Nonetheless, it is possible to find a combination of parameters that produce the negative dispersion necessary for the WLC-SR scheme while producing very low noise at the center of the dip in the gain profile. As an example, we consider next a case where $\Omega_1 = \gamma$, $\Omega_2 = 10^2 \gamma$, $\Omega_3 = 10^{-6} \gamma$, and $\Omega_4 = 10^{-1} \gamma$, whereas the other parameters are the same as those in Fig. 11. The corresponding complex susceptibility is plotted in Fig. 13. As shown in Table I, the steady state of the system in this case is not close to the dark state defined in Eq. (97). Nonetheless, almost all the population is in level $|3\rangle$; therefore, the QN in this case is expected to be very small as confirmed next.

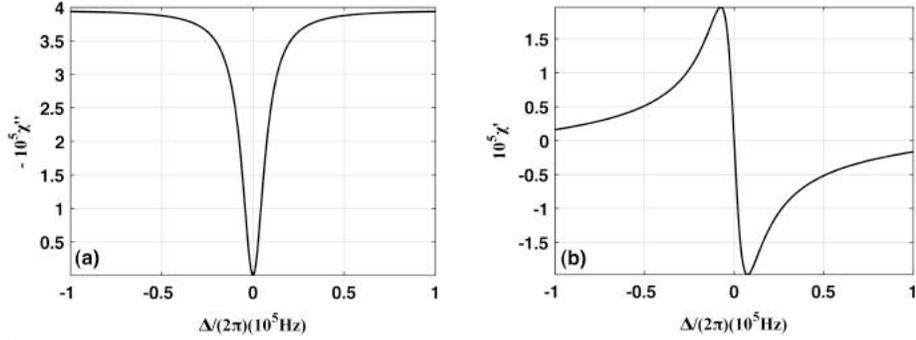

Fig. 13 Plot of (a) imaginary and (b) real part of the complex susceptibility as a function of detuning for the GEIT system from semiclassical calculation. Here $\gamma/(2\pi) = 6 \text{MHz}$, $\Gamma_{41} = \Gamma_{42} = \Gamma_{52} = \Gamma_{53} = \Gamma_{21} = \Gamma_{23} = \gamma/2$, $\delta_1 \simeq \delta_2 \simeq \delta_3 \simeq \delta_4 \simeq 10^3 \gamma$, $\Omega_1 = \gamma$, $\Omega_2 = 10^2 \gamma$, $\Omega_3 = 10^{-6} \gamma$ and $\Omega_4 = 10^{-1} \gamma$.

Table I Comparison of the values of $\rho_{11}$, $\rho_{33}$ and $\rho_{13}$ based on the steady state solution of the density matrix equations and those based on an ideal dark state for the GEIT system using the same parameters used in Fig. 13.

|  | Ideal Dark State Value | Density Matrix Equation |
| --- | --- | --- |
| $\rho_{11}$ | $5.0561 \times 10^{-7}$ | $1.0000 \times 10^{-18}$ |
| $\rho_{33}$ | 1.0000 | 1.0000 |
| $\rho_{13}$ | $-9.9990 \times 10^{-10}$ | $-5.0000 \times 10^{-19}$ |

To apply the ME method to the GEIT system, we start with the equations of motions for the atom-field density operator as in Eq. (2) by treating the pump fields semiclassically and treating only the probe field quantum mechanically. The Hamiltonian for the atom-field system in the interaction picture after adding the complex terms is as follows:

$$\tilde{\mathcal{H}}' = \hbar\left(\delta_2 - \delta_1 - \frac{i}{2}\Gamma_2\right)|2\rangle\langle 2| + \hbar(\delta_2 + \delta_4 - \delta_1 - \delta_3)|3\rangle\langle 3| - \hbar\left(\delta_1 + \frac{i}{2}\Gamma_4\right)|4\rangle\langle 4|$$
$$-\hbar\left(\delta_1 + \delta_3 - \delta_2 + \frac{i}{2}\Gamma_5\right)|5\rangle\langle 5| + \hbar g(|5\rangle\langle 2|a + a^\dagger|2\rangle\langle 5|) - \frac{1}{2}\hbar\Omega_1(|4\rangle\langle 1| + |1\rangle\langle 4|) \quad (101)$$
$$-\frac{1}{2}\hbar\Omega_2(|4\rangle\langle 2| + |2\rangle\langle 4|) - \frac{1}{2}\hbar\Omega_4(|5\rangle\langle 3| + |3\rangle\langle 5|).$$

The equation of motion for the density matrix elements can now be derived from Eq. (2) in the same way as we derived Eqs. (71)–(79), for example. Since the number of equations in this case is rather large (25), we choose to show below only two of the equations that illustrate the fact that elements from adjacent manifolds are coupled, just as in previous cases,

$$\dot{\rho}_{2n+1,2n'+1} = -\Gamma_2\rho_{2n+1,2n'+1} + \Gamma_{42}\rho_{4n+1,4n'+1} + \Gamma_{52}\rho_{5n+1,5n'+1} - \frac{i}{\hbar}(\tilde{\mathcal{H}}'_{2n+1,5n}\rho_{5n,2n'+1} - \rho_{2n+1,5n'}\tilde{\mathcal{H}}'^{*}_{5n',2n'+1}$$
$$+ \tilde{\mathcal{H}}'_{2n+1,4n+1}\rho_{4n+1,2n'+1} - \rho_{2n+1,4n'+1}\tilde{\mathcal{H}}'^{*}_{4n'+1,2n'+1}), \quad (102)$$

$$\dot{\rho}_{3n,3n'} = \Gamma_{53}\rho_{5n,5n'} + \Gamma_{23}\rho_{2n,2n'} - \frac{i}{\hbar}(\tilde{\mathcal{H}}'_{3n,5n}\rho_{5n,3n'} - \rho_{3n,5n'}\tilde{\mathcal{H}}'^{*}_{5n',3n'}). \quad (103)$$

In Eq. (102), $\rho_{5n+1,5n'+1}$ belongs to the ($n$+1) manifold, whereas in Eq. (103), $\rho_{2n,2n'}$ belongs to the ($n$ – 1) manifold. Similar to the approximation we used earlier, we rewrite these terms as

$$\rho_{5n+1,5n'+1} \simeq (1 - \rho_{11}^{(0)} - \rho_{33}^{(0)} - \rho_{44}^{(0)})\tilde{\rho}_{n+1,n'+1} - \rho_{2n+1,2n'+1}, \quad (104)$$

$$\rho_{2n,2n'} \simeq (1 - \rho_{11}^{(0)} - \rho_{44}^{(0)} - \rho_{55}^{(0)})\tilde{\rho}_{nn'} - \rho_{3n,3n'}. \quad (105)$$

Due to the complexity of the system, we cannot get an analytical solution to the 25 equations for the elements of the $n$ manifold. The numerical result is similar to the form of Eq. (9) in Sec. III.A. We have shown in Appendix B 4 that the susceptibilities calculated using the ME model and the semiclassical model essentially agree. For the center detuning $\Delta = 0$, we have $\mathcal{A} \approx 1.3394 \times 10^{-20} g^2$, $\mathcal{B} \approx 1.1627 \times 10^{-22} g^2$. From Eq. (21), we can calculate the noise in this case,

$$\langle \Delta X_\theta^2 \rangle_{ME,noise} = (e^{Gain} - 1)\frac{\mathcal{A} + \mathcal{B}}{4(\mathcal{A} - \mathcal{B})} \approx (e^{Gain} - 1) \times 0.254, \quad (106)$$

where $Gain = (\mathcal{A} - \mathcal{B})T$. If the SC-CM model were used, the noise would be

$$\left\langle \Delta X_\theta^2 \right\rangle_{SC-CM,noise} = \frac{1}{4}\left(e^{Gain} - 1\right), \tag{107}$$

as expected from Eq. (26). If we define

$$\eta = \left\langle \Delta X_\theta^2 \right\rangle_{ME,noise} / \left\langle \Delta X_\theta^2 \right\rangle_{SC-CM,noise} - 1, \tag{108}$$

we find in this case $\eta \approx 1.75\%$, which means that the noise from the GEIT system is 1.75% larger than that from the SC-CM. This difference is not due to numerical errors. In Fig. 14, we show the values of $\eta$ at detunings different from zero. We can see that the noise calculated from the ME and from the SC-CM model agree better than for zero detuning. Thus, the SC-CM model is essentially valid for this choice of parameters of the GEIT system. In Ref. [15], we have used both the SC-CM and the ME approaches to determine the QN-limited enhancement in the sensitivity-bandwidth product. For the case where $\eta = 1.75\%$, the corresponding difference in the sensitivity of the WLC-SR scheme is 0.2%. The factor of enhancement for this case is 16.55 [15]. In Ref. [15], we also showed a different set of parameters ($\Omega_1 = \gamma$, $\Omega_2 = 10^2\gamma$, $\Omega_3 = 10^{-6}\gamma$, $\Omega_4 = 10^{-1}\gamma$, $\delta_1 \simeq \delta_2 \simeq \delta_3 \simeq \delta_4 \simeq 10^3\gamma$, $\Gamma_{41} = \Gamma_{42} = \Gamma_{52} = \Gamma_{53} = \gamma/2$, and $\Gamma_{21} = \Gamma_{23} \approx 2.02 \times 10^{-3}\gamma$) that produces an even higher factor (17.66) of enhancement. In Fig. 15, we show the values of $\eta$ in this case. We find that the prediction of the SC-CM differs significantly from that of the ME model, especially at the center detuning where $\eta$ is on the order of $10^6$. Thus one must use the ME approach to calculate the QN. Comparing the two cases above, we find that the population in the upper level $|4\rangle$ differs by a factor of $10^2$ whereas that in level $|5\rangle$ remains about the same. Therefore, we conclude that the ME model and the SC-CM become more significantly different as the excitation in level $|4\rangle$ increases.

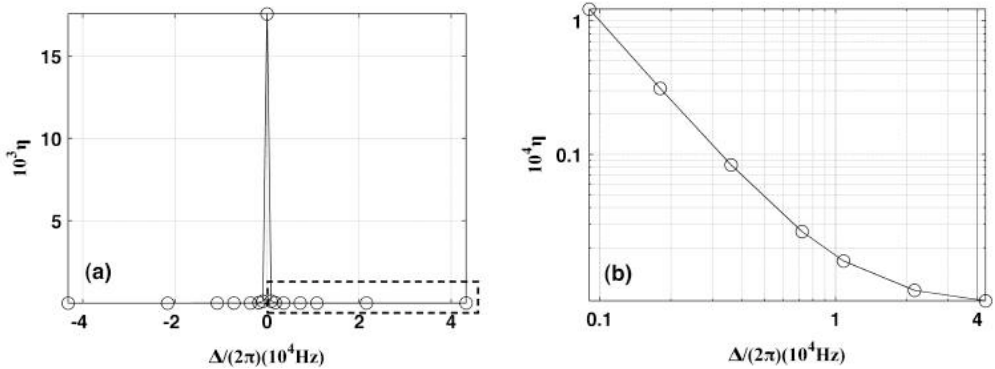

Fig. 14 Plot of $\eta$ as a function of detuning $\Delta/(2\pi)$ when $\gamma/(2\pi) = 6\text{MHz}$, $\Gamma_{41} = \Gamma_{42} = \Gamma_{52} = \Gamma_{53} = \Gamma_{21} = \Gamma_{23} = \gamma/2$, $\delta_1 \approx \delta_2 \approx \delta_3 \approx \delta_4 \approx 10^3\gamma$, $\Omega_1 = \gamma$, $\Omega_2 = 10^2\gamma$, $\Omega_3 = 10^{-6}\gamma$, and $\Omega_4 = 10^{-1}\gamma$. (a) Data points at detunings $\Delta/(2\pi) = 0$, $\pm 9 \times 10^2$, $\pm 1.8 \times 10^3$, $\pm 3.6 \times 10^3$, $\pm 7.2 \times 10^3$, $\pm 1.08 \times 10^4$, $\pm 2.16 \times 10^4$, and $\pm 4.32 \times 10^4$ Hz are shown in asterisks; (b) zoom in and plot the part of positive detunings only, which is shown in the dashed box in (a). The difference as high as $10^{-2}$ is not due to numerical errors.

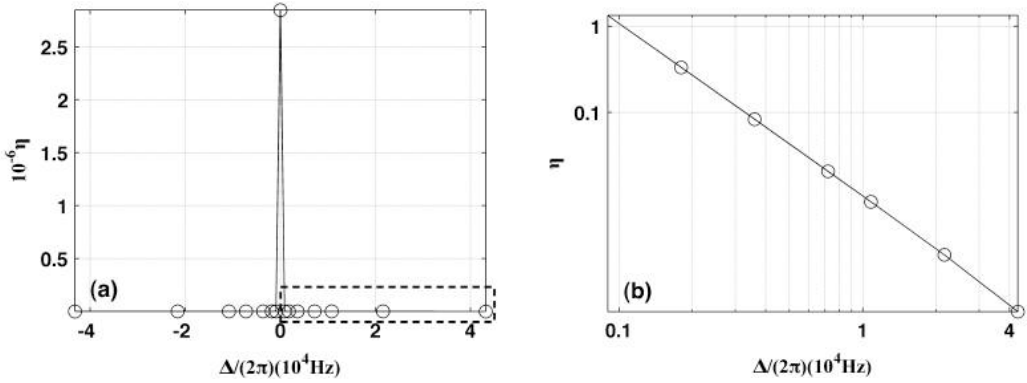

Fig. 15 Plot of $\eta$ as a function of detuning $\Delta/(2\pi)$ when $\gamma/(2\pi) = 6\text{MHz}$, $\Gamma_{41} = \Gamma_{42} = \Gamma_{52} = \Gamma_{53} = \gamma/2$, $\Gamma_{21} = \Gamma_{23} \approx 2.02 \times 10^{-3}\gamma$, $\delta_1 \approx \delta_2 \approx \delta_3 \approx \delta_4 \approx 10^3\gamma$, $\Omega_1 = \gamma$, $\Omega_2 = 10^2\gamma$, $\Omega_3 = 10^{-6}\gamma$, and $\Omega_4 = 10^{-1}\gamma$.

## VII. CONCLUSION

We have used the ME approach to derive explicitly the QN for a field interacting with four different kinds of resonant or near-resonant atomic systems, and compared the results with those predicted by the SC-CM [1,2]. In developing this model, we find that, for all systems other than a simple two level transition, it is necessary to make a steady-state approximation for the evolution of the density matrix for the combined system of atoms and the quantized probe field in order to decouple different manifolds corresponding to different numbers of photons. This approximation enables us to determine explicitly the reduced density-matrix equation of motion for the quantized probe field. We verify the validity of this approximation by showing that the atomic susceptibility yielded by the ME model agrees very well with the same derived via semiclassical analysis. While comparing the QN calculated by the ME model with the prediction of the SC-CM, we find that in some cases, there are significant differences. Specifically, we have shown that in a four-level system with an absorption dip on top of a broad gain peak and perfect transparency at the center, the net QN at the center has a large finite value in sharp contrast with the SC-CM, which predicts zero QN at the center. This catastrophic breakdown of the Caves model implies that it is incorrect to assume the SC-CM to be a close approximation of the more exact result. We find that the SC-CM model result varies significantly from the ME result even for a two-level system under most conditions. We also show a special case where the result of the ME model is in agreement with that of the SC-CM. This case is a $\Lambda$-type EIT system, for which the QN at the center detuning is zero while the system is in the dark state. Finally, we describe a GEIT system, which has a negative dispersion and a similar transmission profile but much lower QN around the center compared to the four-level atomic system mentioned above. In this case, we find that, for some set of parameters, the QN as predicted by the SC-CM agrees closely with the ME model. However, we also find that for some other set of parameters, the SC-CM model disagrees strongly with the ME model. In general, the QN predicted by the SC-CM is less than or equal to that predicted by the ME model. However, we do not find a general rule that can be applied to determine when the application of the SC-CM is expected to be a good approximation of the more exact result. Therefore, one must always make use of the ME approach when dealing with resonant or near-resonant atomic systems. The technique presented in this paper would enable accurate evaluation of the QN in many systems of interest in precision metrology.

## ACKNOWLEDGMENTS

This work was supported by DARPA through the slow light program under Grant No. FA9550-07-C-0030 and by AFOSR under Grant No. FA9550-10-1-0228 and No. FA9550-09-1-0652.

## APPENDIX A

Here we show the detailed steps of the master equation approach for the two-level atomic system. The Hamiltonian for the atom-field system, under the rotating-wave approximation, can be written as

$$\mathcal{H}_T = \hbar \upsilon a^\dagger a + \hbar \omega_a |a\rangle\langle a| + \hbar \omega_b |b\rangle\langle b| + \hbar g(\sigma_+ a + a^\dagger \sigma_-). \tag{A1}$$

We then use a rotating wave transformation to the interaction picture where the state vector is transformed as

$$|\psi_I(t)\rangle = U_0^\dagger(t)|\psi(t)\rangle, \text{where } U_0(t) = \exp(-i\mathcal{H}_0 t / \hbar) \tag{A2}$$

with

$$\mathcal{H}_0 = \hbar v a^\dagger a + \hbar \omega_a |a\rangle\langle a| + \hbar(\omega_a - v)|b\rangle\langle b|, \tag{A3}$$

$$\mathcal{H} = \mathcal{H}_T - \mathcal{H}_0 = \hbar\Delta|b\rangle\langle b| + \hbar g(\sigma_+ a + a^\dagger \sigma_-). \tag{A4}$$

where $\Delta = v - \omega$. As a result the Hamiltonian after rotating-wave transformation is then

$$\tilde{\mathcal{H}} = U_0^\dagger \mathcal{H} U_0 = \hbar\Delta|b\rangle\langle b| + \hbar g(\sigma_+ a + a^\dagger \sigma_-), \tag{A5}$$

and the density operator becomes $\rho = U_0^\dagger \varrho U_0$. The equation of motion given by Eq. (1) then becomes

$$\dot{\rho}_{a-f} = -\frac{i}{\hbar}\left[\tilde{\mathcal{H}}, \rho_{a-f}\right] + \dot{\rho}_{a-f,R} \tag{A6}$$

As described in Sec. II, we add complex terms in the atomic Hamiltonian to take into account the decay and dephasing of the atom, and we then have the Hamiltonian,

$$\tilde{\mathcal{H}}' = \tilde{\mathcal{H}} - \frac{i}{2}\hbar\gamma_a|a\rangle\langle a| - \frac{i}{2}\hbar\gamma_{op}|b\rangle\langle b|. \tag{A7}$$

Therefore the equation of motion can be written as

$$\dot{\rho}_{a-f} = -\frac{i}{\hbar}\left(\tilde{\mathcal{H}}'\rho_{a-f} - \rho_{a-f}\tilde{\mathcal{H}}'^*\right) + \dot{\rho}_{a-f,Source}. \tag{A8}$$

It can be shown that the matrix elements of the Hamiltonian $\tilde{\mathcal{H}}'_{\alpha n,\beta n'} = \langle\alpha,n|\tilde{\mathcal{H}}'|\beta,n'\rangle$ are as follows:

$$\tilde{\mathcal{H}}'_{an,an} = -i\hbar\gamma_a/2, \tilde{\mathcal{H}}'_{bn+1,bn+1} = \hbar\left(\Delta - i\gamma_{op}/2\right), \tilde{\mathcal{H}}'_{an,bn+1} = \hbar g\sqrt{n+1} = \tilde{\mathcal{H}}'_{bn+1,an} = \tilde{\mathcal{H}}'^*_{bn+1,an}, \tag{A9}$$

and zero otherwise. The equations of motion for the density matrix $\rho_{\alpha n,\beta n'} = \langle\alpha,n|\rho_{atom-field}|\beta,n'\rangle$ can be derived from Eq. (A8), i.e.,

$$\dot{\rho}_{an,an'} = \gamma_{op}\rho_{bn,bn'} - \gamma_a\rho_{an,an'} - \frac{i}{\hbar}(\tilde{\mathcal{H}}'_{an,bn+1}\rho_{bn+1,an'} - \rho_{an,bn'+1}\tilde{\mathcal{H}}'^*_{bn'+1,an'}), \tag{A10}$$

$$\dot{\rho}_{an,bn'+1} = -(\gamma_{ab} - i\Delta)\rho_{an,bn'+1} - \frac{i}{\hbar}(\tilde{\mathcal{H}}'_{an,bn+1}\rho_{bn+1,bn'+1} - \rho_{an,an'}\tilde{\mathcal{H}}'^*_{an',bn'+1}), \tag{A11}$$

$$\dot{\rho}_{bn+1,an'} = -(\gamma_{ab} + i\Delta)\rho_{bn+1,an'} - \frac{i}{\hbar}(\tilde{\mathcal{H}}'_{bn+1,an}\rho_{an,an'} - \rho_{bn+1,bn'+1}\tilde{\mathcal{H}}'^*_{bn'+1,an'}), \tag{A12}$$

$$\dot{\rho}_{bn+1,bn'+1} = \gamma_a\rho_{an+1,an'+1} - \gamma_{op}\rho_{bn+1,bn'+1} - \frac{i}{\hbar}(\tilde{\mathcal{H}}'_{bn+1,an}\rho_{an,bn'+1} - \rho_{bn+1,an'}\tilde{\mathcal{H}}'^*_{an',bn'+1}), \tag{A13}$$

where $\gamma_{ab} = (\gamma_a + \gamma_{op})/2$. When $\Delta = 0$, these equations are essentially the same as Eqs. (11.1.5a)–(11.1.5g) in Ref. [16] when the differences in the various pumping and decay rates are taken into account.

Using the relation that $\rho_{an,an'} + \rho_{bn,bn'} = \tilde{\rho}_{nn'}$ and $\rho_{an+1,an'+1} + \rho_{bn+1,bn'+1} = \tilde{\rho}_{n+1,n'+1}$, we rewrite Eqs. (A10)–(A13) as follows:

$$\dot{\rho}_{an,an'} = \gamma_{op}\left(\tilde{\rho}_{nn'} - \rho_{an,an'}\right) - \gamma_a \rho_{an,an'} - \frac{i}{\hbar}(\tilde{\mathcal{H}}'_{an,bn+1}\rho_{bn+1,an'} - \rho_{an,bn'+1}\tilde{\mathcal{H}}'^*_{bn'+1,an'}), \tag{A14}$$

$$\dot{\rho}_{bn+1,bn'+1} = \gamma_a\left(\tilde{\rho}_{n+1,n'+1} - \rho_{bn+1,bn'+1}\right) - \gamma_{op}\rho_{bn+1,bn'+1} - \frac{i}{\hbar}(\tilde{\mathcal{H}}'_{bn+1,an}\rho_{an,bn'+1} - \rho_{bn+1,an'}\tilde{\mathcal{H}}'^*_{an',bn'+1}). \tag{A15}$$

The set of equations can be solved by rewriting it in the matrix form [16]

$$\dot{R} = -MR + A, \tag{A16}$$

where

$$M = \begin{pmatrix} 2\gamma_{ab} & -i\tilde{\mathcal{H}}'^*_{bn'+1,an'}/\hbar & i\tilde{\mathcal{H}}'_{an,bn+1}/\hbar & 0 \\ -i\tilde{\mathcal{H}}'^*_{an',bn'+1}/\hbar & \gamma_{ab} - i\Delta & 0 & i\tilde{\mathcal{H}}'_{an,bn+1}/\hbar \\ i\tilde{\mathcal{H}}'_{bn+1,an}/\hbar & 0 & \gamma_{ab} + i\Delta & -i\tilde{\mathcal{H}}'^*_{bn'+1,an'}/\hbar \\ 0 & i\tilde{\mathcal{H}}'_{bn+1,an}/\hbar & -i\tilde{\mathcal{H}}'^*_{an',bn'+1}/\hbar & 2\gamma_{ab} \end{pmatrix}, \tag{A17}$$

$$R = \left(\rho_{an,an'}, \rho_{an,bn'+1}, \rho_{bn+1,an'}, \rho_{bn+1,bn'+1}\right)^T, A = \left(\gamma_{op}\tilde{\rho}_{nn'}, 0, 0, \gamma_a\tilde{\rho}_{n+1,n'+1}\right)^T. \tag{A18}$$

In the adiabatic limit, we assume that $R$ varies slowly compared to $|M|$. In this limit, we can set $\dot{R} \simeq 0$, to get (Ref. [16], Chapter 11),

$$R = M^{-1}A. \tag{A19}$$

The results for the density-matrix elements are

$$\rho_{an,bn'+1} = ig\frac{2\gamma_{ab}(\Delta + i\gamma_{ab})\left(\sqrt{n'+1}\gamma_{op}\tilde{\rho}_{nn'} - \sqrt{n+1}\gamma_a\tilde{\rho}_{n+1,n'+1}\right) - ig^2(n'-n)\left(\sqrt{n'+1}\gamma_{op}\tilde{\rho}_{nn'} + \sqrt{n+1}\gamma_a\tilde{\rho}_{n+1,n'+1}\right)}{g^4(n-n')^2 + 4g^2(n+n'+2)\gamma_{ab}^2 + 4\gamma_{ab}^2\left(\gamma_{ab}^2 + \Delta^2\right)}, \tag{A20}$$

$$\rho_{bn+1,an'} = ig\frac{2\gamma_{ab}(\Delta - i\gamma_{ab})\left(\sqrt{n+1}\gamma_{op}\tilde{\rho}_{nn'} - \sqrt{n'+1}\gamma_a\tilde{\rho}_{n+1,n'+1}\right) - ig^2(n-n')\left(\sqrt{n+1}\gamma_{op}\tilde{\rho}_{nn'} + \sqrt{n'+1}\gamma_a\tilde{\rho}_{n+1,n'+1}\right)}{g^4(n-n')^2 + 4g^2(n+n'+2)\gamma_{ab}^2 + 4\gamma_{ab}^2\left(\gamma_{ab}^2 + \Delta^2\right)}. \tag{A21}$$

In the linear regime where $g$ is very small, the results above can be simplified as

$$\rho_{an,bn'+1} = \frac{ig\left(\sqrt{n'+1}\gamma_{op}\tilde{\rho}_{nn'} - \sqrt{n+1}\gamma_a\tilde{\rho}_{n+1,n'+1}\right)}{2\gamma_{ab}(\gamma_{ab} + i\Delta)}, \quad \rho_{bn+1,an'} = \frac{-ig\left(\sqrt{n+1}\gamma_{op}\tilde{\rho}_{nn'} - \sqrt{n'+1}\gamma_a\tilde{\rho}_{n+1,n'+1}\right)}{2\gamma_{ab}(\gamma_{ab} - i\Delta)}. \tag{A22}$$

Using

$$\dot{\tilde{\rho}} = -\frac{i}{\hbar}Tr_{atom}\left(\tilde{\mathcal{H}}'\rho_{a-f} - \rho_{a-f}\tilde{\mathcal{H}}'^*\right), \tag{A23}$$

we then get the equation of motion for the reduced density matrix of the field given by

$$\dot{\rho}_{nn'} = -\frac{i}{\hbar}\left(\tilde{\mathcal{H}}'_{an,bn+1}\rho_{bn+1,an'} - \rho_{an,bn'+1}\tilde{\mathcal{H}}'^{*}_{bn'+1,an'} + \tilde{\mathcal{H}}'_{bn,an-1}\rho_{an-1,bn'} - \rho_{bn,an'-1}\tilde{\mathcal{H}}'^{*}_{an'-1,bn'}\right), \tag{A24}$$

It is easy to see that the complex parts of $\tilde{\mathcal{H}}'$ cancel out so that this can be expressed as:

$$\dot{\rho}_{nn'} = -\frac{i}{\hbar}\left(\tilde{\mathcal{H}}_{an,bn+1}\rho_{bn+1,an'} - \rho_{an,bn'+1}\tilde{\mathcal{H}}_{bn'+1,an'} + \tilde{\mathcal{H}}_{bn,an-1}\rho_{an-1,bn'} - \rho_{bn,an'-1}\tilde{\mathcal{H}}_{an'-1,bn'}\right). \tag{A25}$$

For $\Delta = 0$, these equations are essentially the same as Eq. (11.1.3d) in Ref. [16] when differences in notations and various pumping and decay rates are taken into account.

## APPENDIX B

In this Appendix, we show the derivations of susceptibilities for the atomic systems we discussed above following both the ME model and the semiclassical approach and compare the results.

1. Two-level atomic system

Solving the semiclassical equations of motion for the density matrix of the atom, we can get

$$\rho_{ab} = \frac{\Omega(\gamma_a - \gamma_{op})}{4\gamma_{ab}^2 + 4\Delta^2 + 2\Omega^2}\frac{\Delta}{\gamma_{ab}} - i\frac{\Omega(\gamma_a - \gamma_{op})}{4\gamma_{ab}^2 + 4\Delta^2 + 2\Omega^2}. \tag{B1}$$

Then,

$$\chi = -\frac{2\mathcal{P}}{\mathcal{E}_0}n\rho_{ab} = -\frac{2\hbar\Omega}{\mathcal{E}_0^2}n\rho_{ab} = \chi' + i\chi'', \tag{B2}$$

where $\mathcal{P} = e\langle a|r|b\rangle$ is the electric dipole transition matrix element, and $\Omega = \mathcal{P}\mathcal{E}_0/\hbar$ is the Rabi frequency. Therefore,

$$\chi'' = \frac{2\hbar\Omega}{\mathcal{E}_0^2}n\frac{\Omega(\gamma_a - \gamma_{op})}{4\gamma_{ab}^2 + 4\Delta^2 + 2\Omega^2}, \tag{B3}$$

$$\chi' = -\frac{\Delta}{\gamma_{ab}}\chi''. \tag{B4}$$

For simplicity, we now limit our discussion to the pure attenuation case where $\gamma_a = \gamma$ and $\gamma_{op} = 0$. Therefore,

$$\chi'' = \frac{2\hbar\Omega}{\mathcal{E}_0^2} n \frac{\Omega\gamma}{\gamma^2 + 4\Delta^2 + 2\Omega^2}, \tag{B5}$$

$$\chi' = -\frac{2\Delta}{\gamma}\chi''. \tag{B6}$$

Since we limit our quantum results to the linear regime where $\Omega \to 0$, we make the same approximation here. Then the total gain for a field traveling a distance $L$ is

$$G_{SC} = -\frac{1}{2} 4\pi\chi'' kL = -\frac{4\pi\hbar\Omega^2}{\mathcal{E}_0^2} n \frac{\gamma}{\gamma^2 + 4\Delta^2} kL \tag{B7}$$

Notice that $g = -\mathcal{P}\mathcal{E}/\hbar$, where $\mathcal{E} = (2\pi\hbar\nu/V)^{1/2}$, whereas $\Omega = \mathcal{P}\mathcal{E}_0/\hbar$ with $\mathcal{E}_0^2 V/(8\pi) = \hbar\nu$, i.e., $\mathcal{E}_0 = (8\pi\hbar\nu/V)^{1/2}$. Therefore $g = \Omega/2$, and we get

$$G_{SC} = -nV \frac{2g^2\gamma}{\gamma^2 + 4\Delta^2} \frac{L}{c}. \tag{B8}$$

The total phase shift the field experiences is then

$$\theta_{SC} = \frac{1}{2} 4\pi\chi' kL = \frac{2\Delta}{\gamma} G_{SC} \tag{B9}$$

On the other hand, from the ME results in Eq. (33), we can get

$$\mathcal{B} = \frac{2g^2\gamma}{\gamma^2 + 4\Delta^2}, \quad \mathcal{N} = -\frac{2g^2\gamma_a}{(\gamma^2 + 4\Delta^2)} \frac{2\Delta}{\gamma} = -\frac{2\Delta}{\gamma}\mathcal{B}, \tag{B10}$$

Therefore, the total gain and phase shift are

$$G_{ME} = -N_{atom}\mathcal{B}t = -nV \frac{2g^2\gamma}{\gamma^2 + 4\Delta^2} \frac{L}{c}, \tag{B11}$$

$$\theta_{ME} = N_{atom}\mathcal{N}t = \frac{2\Delta}{\gamma} G_{ME}, \tag{B12}$$

which are the same as the semiclassical results in Eqs. (B8) and (B9). Thus we have shown that the results of the gain and phase shift that the field experiences in a two-level atomic system computed using the ME method and the semiclassical approach agree with each other.

2. Four-level atomic system

We now compare the susceptibilities of the system considered in Sec. IV with the semiclassical results. Denote the Rabi frequency of the field as $\Omega$. Solving the density matrix equations, we can get

$$\rho_{ab} = \frac{\Omega(\gamma_a - \gamma_{op})(\Delta - i\gamma_{ab})}{2(2\gamma_a + \gamma_{op})(\gamma_{ab}^2 + \Delta^2)}, \quad \rho_{cd} = \frac{\Omega \gamma_a (\Delta + \delta - i\gamma_{cd})}{2(2\gamma_a + \gamma_{op})[\gamma_{cd}^2 + (\Delta + \delta)^2]}, \quad (B13)$$

in the linear regime where $\Omega$ is very small. Then,

$$\chi = -\frac{2\hbar\Omega}{\mathcal{E}_0^2} n(\rho_{ab} + \rho_{cd}) = \chi' + i\chi'', \quad (B14)$$

where the imaginary and real parts of susceptibility are as follows:

$$\chi'' = \frac{2\hbar\Omega^2}{\mathcal{E}_0^2} n \left\{ \frac{(\gamma_a - \gamma_{op})\gamma_{ab}}{2(2\gamma_a + \gamma_{op})(\gamma_{ab}^2 + \Delta^2)} + \frac{\gamma_a \gamma_{cd}}{2(2\gamma_a + \gamma_{op})[\gamma_{cd}^2 + (\Delta + \delta)^2]} \right\}, \quad (B15)$$

$$\chi' = -\frac{2\hbar\Omega^2}{\mathcal{E}_0^2} n \left\{ \frac{(\gamma_a - \gamma_{op})\Delta}{2(2\gamma_a + \gamma_{op})(\gamma_{ab}^2 + \Delta^2)} + \frac{\gamma_a (\Delta + \delta)}{2(2\gamma_a + \gamma_{op})[\gamma_{cd}^2 + (\Delta + \delta)^2]} \right\}, \quad (B16)$$

Then the total gain for a field traveling a distance $L$ is

$$G_{SC} = -\frac{1}{2} 4\pi \chi'' kL = \frac{4\pi \hbar \Omega^2}{\mathcal{E}_0^2} n \left\{ \frac{(\gamma_{op} - \gamma_a)\gamma_{ab}}{2(2\gamma_a + \gamma_{op})(\gamma_{ab}^2 + \Delta^2)} - \frac{\gamma_a \gamma_{cd}}{2(2\gamma_a + \gamma_{op})[\gamma_{cd}^2 + (\Delta + \delta)^2]} \right\} kL, \quad (B17)$$

Using the relation that $g = \Omega_s / 2$, we can write

$$G_{SC} = nVg^2 \left\{ \frac{\gamma_{ab}(\gamma_{op} - \gamma_a)}{(\gamma_{ab}^2 + \Delta^2)(2\gamma_a + \gamma_{op})} - \frac{\gamma_{cd}\gamma_a}{[\gamma_{cd}^2 + (\Delta + \delta)^2](2\gamma_a + \gamma_{op})} \right\} \frac{L}{c}. \quad (B18)$$

Similarly, the total phase shift is

$$\theta_{SC} = \frac{1}{2} 4\pi \chi' kL = nVg^2 \left\{ \frac{(\gamma_{op} - \gamma_a)\Delta}{(2\gamma_a + \gamma_{op})(\gamma_{ab}^2 + \Delta^2)} - \frac{\gamma_a(\Delta + \delta)}{(2\gamma_a + \gamma_{op})[\gamma_{cd}^2 + (\Delta + \delta)^2]} \right\} \frac{L}{c}. \quad (B19)$$

On the other hand, from the ME results in Eq. (58) and (59), we have the total gain and phase shift, respectively, as

$$G_{ME} = N_{atom}(\mathcal{A} - \mathcal{B} - \mathcal{D})t = nVg^2 \left\{ \frac{\gamma_{ab}(\gamma_{op} - \gamma_a)}{(\gamma_{ab}^2 + \Delta^2)(2\gamma_a + \gamma_{op})} - \frac{\gamma_{cd}\gamma_a}{[\gamma_{cd}^2 + (\Delta + \delta)^2](2\gamma_a + \gamma_{op})} \right\} \frac{L}{c}, \quad (B20)$$

$$\theta_{ME} = N_{atom}\mathcal{N}t = nVg^2\left\{\frac{(\gamma_{op}-\gamma_a)\Delta}{(2\gamma_a+\gamma_{op})(\gamma_{ab}^2+\Delta^2)} - \frac{\gamma_a(\Delta+\delta)}{(2\gamma_a+\gamma_{op})\left[\gamma_{cd}^2+(\Delta+\delta)^2\right]}\right\}\frac{L}{c}, \tag{B21}$$

both of which agree with the semiclassical results in Eqs. (B17) and (B18). This validates the approximations we employed in deriving Eqs. (49)–(51).

### 3. Λ-type EIT system

We denote the Rabi frequency of the pump field as $\Omega_p$ and the Rabi frequency of the probe field as $\Omega_s$. We now solve the semiclassical equations of motion for the density matrix to get the result for a general value of the detuning $\Delta$,

$$\rho_{aa} = \frac{8\Omega_s^2\Omega_p^2\Delta^2}{16\Delta^4\Omega_p^2 + 4\gamma^2\Delta^2(\Omega_p^2+\Omega_s^2) + (\Omega_p^2+\Omega_s^2)^3 - 8\Delta^2\Omega_p^2(\Omega_p^2-2\Omega_s^2)}, \tag{B22}$$

$$\rho_{bb} = \frac{8\Omega_p^2\left[4\gamma^2\Delta^2 + 16\Delta^4 + (\Omega_p^2+\Omega_s^2)^2 - 4\Delta^2(2\Omega_p^2-\Omega_s^2)\right]}{16\Delta^4\Omega_p^2 + 4\gamma^2\Delta^2(\Omega_p^2+\Omega_s^2) + (\Omega_p^2+\Omega_s^2)^3 - 8\Delta^2\Omega_p^2(\Omega_p^2-2\Omega_s^2)}, \tag{B23}$$

$$\rho_{cc} = \frac{\Omega_s^2\left[4\gamma^2\Delta^2 + 4\Delta^2\Omega_p^2 + (\Omega_p^2+\Omega_s^2)^2\right]}{16\Delta^4\Omega_p^2 + 4\gamma^2\Delta^2(\Omega_p^2+\Omega_s^2) + (\Omega_p^2+\Omega_s^2)^3 - 8\Delta^2\Omega_p^2(\Omega_p^2-2\Omega_s^2)}, \tag{B24}$$

$$\rho_{ab} = \rho_{ba}^* = \frac{2\Delta\Omega_p^2\Omega_s(-2i\gamma\Delta + 4\Delta^2 - \Omega_p^2 - \Omega_s^2)}{16\Delta^4\Omega_p^2 + 4\gamma^2\Delta^2(\Omega_p^2+\Omega_s^2) + (\Omega_p^2+\Omega_s^2)^3 - 8\Delta^2\Omega_p^2(\Omega_p^2-2\Omega_s^2)}, \tag{B25}$$

$$\rho_{ac} = \rho_{ca}^* = \frac{2\Delta\Omega_p\Omega_s^2(-2i\gamma\Delta + \Omega_p^2 + \Omega_s^2)}{16\Delta^4\Omega_p^2 + 4\gamma^2\Delta^2(\Omega_p^2+\Omega_s^2) + (\Omega_p^2+\Omega_s^2)^3 - 8\Delta^2\Omega_p^2(\Omega_p^2-2\Omega_s^2)}, \tag{B26}$$

$$\rho_{bc} = \rho_{cb}^* = \frac{\Omega_p\Omega_s\left[-4\Delta^2\Omega_p^2 - 2i\gamma\Delta(\Omega_p^2+\Omega_s^2) + (\Omega_p^2+\Omega_s^2)^2\right]}{16\Delta^4\Omega_p^2 + 4\gamma^2\Delta^2(\Omega_p^2+\Omega_s^2) + (\Omega_p^2+\Omega_s^2)^3 - 8\Delta^2\Omega_p^2(\Omega_p^2-2\Omega_s^2)}. \tag{B27}$$

Using Eq. (B25) in the limit $\Omega_p \gg \Omega_s$, we can calculate the real and imaginary parts of the susceptibility,

$$\chi'' = \frac{2\hbar\Omega_s}{\mathcal{E}_0^2}n\frac{4\Omega_s\gamma\Delta^2}{16\Delta^4 + 4\gamma^2\Delta^2 + \Omega_p^2 - 8\Delta^2\Omega_p^2}, \tag{B28}$$

$$\chi' = -\frac{4\Delta^2 - \Omega_p^2}{2\gamma\Delta}\chi''. \tag{B29}$$

Then the total gain for a field traveling a distance $L$ is

$$G_{SC} = -\frac{1}{2}4\pi\chi''kL = -\frac{4\pi\hbar\Omega_s^2}{\mathcal{E}_0^2}n\frac{4\gamma\Delta^2}{16\Delta^4 + 4\gamma^2\Delta^2 + \Omega_p^4 - 8\Delta^2\Omega_p^2}kL. \tag{B30}$$

Using the relation that $g = \Omega_s/2$, we can write

$$G_{SC} = -nV\frac{g^2\Delta^2\gamma/2}{\gamma^2\Delta^2/4 + \left(\Delta^2 - \Omega_p^2/4\right)^2}\frac{L}{c}. \tag{B31}$$

The total phase shift the field experiences is then

$$\theta_{SC} = \frac{1}{2}\chi'kL = \frac{4\Delta^2 - \Omega_p^2}{2\gamma\Delta}G_{SC}. \tag{B32}$$

On the other hand, from the ME results in Eq. (84), we can get

$$\mathcal{B} = \frac{g^2\Delta^2\gamma/2}{\gamma^2\Delta^2/4 + \left(\Delta^2 - \Omega_p^2/4\right)^2}, \quad \mathcal{N} = -\frac{4\Delta^2 - \Omega_p^2}{2\gamma\Delta}\mathcal{B}. \tag{B33}$$

Therefore, the total gain and phase shift are

$$G_{ME} = -N_{atom}\mathcal{B}t = -nV\frac{g^2\Delta^2\gamma/2}{\gamma^2\Delta^2/4 + \left(\Delta^2 - \Omega_p^2/4\right)^2}\frac{L}{c}, \tag{B34}$$

$$\theta_{ME} = N_{atom}\mathcal{N}t = \frac{4\Delta^2 - \Omega_p^2}{2\gamma\Delta}G_{ME},, \tag{B35}$$

which agree with the semiclassical results in Eqs. (B31) and (B32). Again, this justifies the approximations made in arriving at Eq. (81).

### 4. Five-level GEIT system

In the five-level GEIT system, we get a numerical solution following the ME approach as shown in Sec. VI. Using the relationship between the ME result and the semiclassical susceptibility as shown in the previous sections, we can convert the ME results for different detunings to the values of the complex susceptibility. In Fig. 16, we show the real and imaginary parts of the susceptibilities calculated using the ME model and the semiclassical approach. The agreement of the two models is essentially exact, and the residual fractional difference, being on the order of $10^{-9}$, is most likely due to inherent inaccuracies of numerical computation.

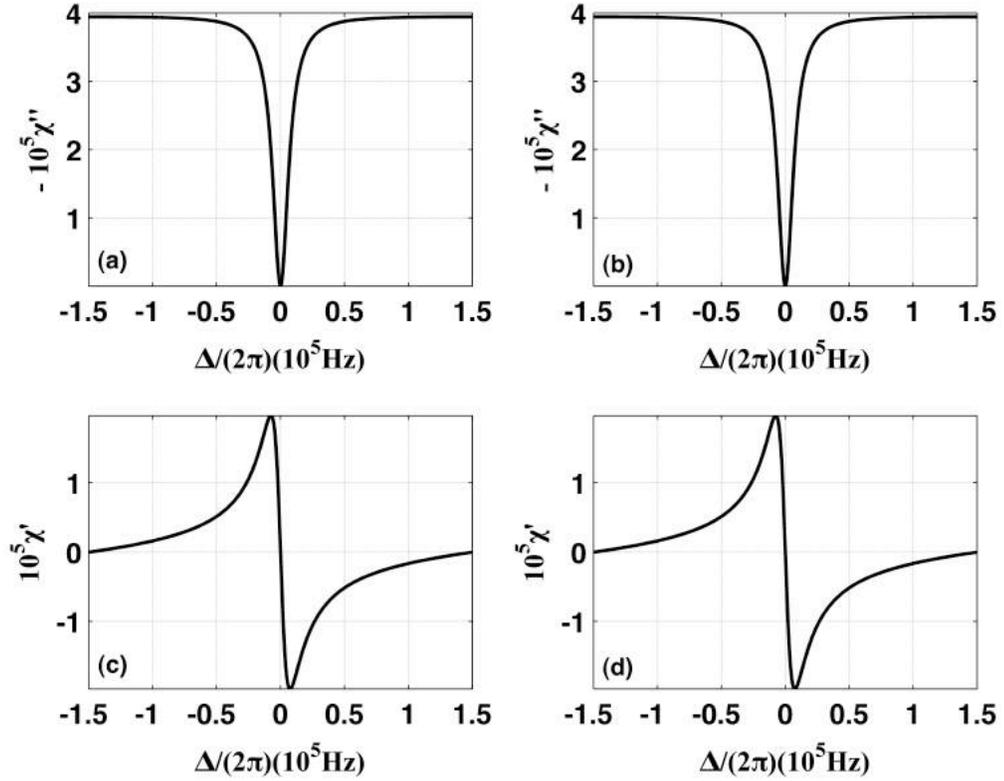

Fig. 16 The left column shows plots of imaginary (a) and real (c) part of the susceptibilities as a function of detuning $\Delta/(2\pi)$ using the ME model. The right column shows plots of imaginary (b) and real (d) part of the susceptibilities using the semiclassical approach. The agreement of the two models is essentially exact, and the residual fractional difference, being of the order of $10^{-9}$, is most likely due to inherent inaccuracies of numerical computation.